\documentclass[%
 reprint,
nofootinbib,
fn2-in-1,
showkeys,
 amsmath,amssymb,
 aps,
]{revtex4-2}

\usepackage{graphicx}
\usepackage{xcolor}
\usepackage{bm,url,physics}

\usepackage{amsmath,amssymb,ascmac,comment}

\begin{document}

\preprint{}

\title{Cooling of Isolated Neutron Stars with Hyperon-mixed Kaon-Condensation Matter
}

\author{Bhavnesh Bhat}
\affiliation{Jodrell Bank Centre for Astrophysics, Department of Physics and Astronomy, The University of Manchester, Manchester M13 9PL, UK}
\affiliation{Indian Institute of Technology, Roorkee, India}

\author{Akira Dohi}
\affiliation{Astrophysical Big Bang Laboratory (ABBL), Cluster for Pioneering Research, RIKEN, Wako, Saitama 351-0198, Japan}
\affiliation{Interdisciplinary Theoretical and Mathematical Sciences (iTHEMS) Center, RIKEN, Wako, Saitama 351-0198, Japan}

\author{Takumi Muto}
\affiliation{Department of Physics, Chiba Institute of Technology, 2-1-1 Shibazono, Narashino, Chiba 275-0023,
Japan}

\author{Tsuneo Noda}
\affiliation{Department of Education and Creation Engineering, Kurume Institute of Technology, Kurume, Fukuoka 830-0052, Japan}

\date{\today}

\begin{abstract}

We investigate the thermal evolution of isolated neutron stars containing hyperon--mixed kaon--condensed matter, focusing on the role of proton superconductivity. The equation of state utilized for cooling calculation is based upon the minimal relativistic mean--field framework supplemented by chiral SU(3) dynamics for kaon condensation with an additional component on the three-baryon force, which ensures stiffness at high densities enough to meet astrophysical constraints on neutron-star masses and radii. We show that the nucleonic direct Urca processes operate at relatively low stellar masses
($M \gtrsim 1.3\,M_\odot$), erasing any observable signature of strangeness in the absence of superfluidity. However, if the proton $^1{\rm S}_0$ superconductivity works, because of suppression of fast neutrino cooling processes, the cooling scenario could become relevant with the strangeness, depending on the density regions of the pairing gap. In particular, if the proton superconductivity is so strong in high-density regions ($T_{c,p}\sim10^{10}~{\rm K}$), the nucleon and hyperon direct Urca processes shut down, which makes the kaon-induced Urca processes dominant in massive neutron stars. This scenario is in good agreement with several cold isolated neutron stars identified recently, such as the Vela Jr., PSR J0205$+$6449, and PSR B2334$+$61. Hence, we suggest that strong proton superconductivity can render kaon condensation observationally visible through cold neutron-star observations, providing a potential signature of strangeness in dense matter.


\end{abstract}

\maketitle


\newcommand{\apjs}{Astrophys. J. Suppl. }
\newcommand{\apjl}{Astrophys. J. Lett. }
\newcommand{\pasj}{Publ. Astron. Soc. Japan. }
\newcommand{\pasa}{Publ. Astron. Soc. Australia. }
\newcommand{\physrep}{Phys. Rep. }
\newcommand{\ptp}{Prog. Theor. Phys. }
\newcommand{\ptps}{Prog. Theor. Phys. Suppl. }
\newcommand{\ptep}{Prog. Theor. Exp. Phys. }
\newcommand{\AIP}{AIP Conf. Proc. }
\newcommand{\aap}{Astron. Astrophys. }
\newcommand{\ssr}{Space Sci. Rev. }
\newcommand{\sci}{Science }
\newcommand{\nar}{New Astron. Rev. }

\newcommand{\araa}{Ann. Rev. Astron. Astrophy. }
\newcommand{\mnras}{Mon. Not. Roy. Astron. Soc. }

\newcommand{\nphysa}{Nucl. Phys. A}

\newcommand{\jcap}{JCAP}

\newcommand{\memsai}{Memorie della Soc. Astron. Ital.}


\section{Introduction}

Observations of neutron stars (NSs), which are the densest stars in the Universe, give us information about their interior. The most conventional observations are the NS mass and radius, which can probe thermodynamic pressure, i.e., the stiffness of high-density matter. From the stiffness, one can \textit{indirectly} examine the compositions inside NSs, e.g., the fraction of hyperons, mesons, quarks, and possible other exotic particles. However, since it is the total pressure that includes all possible particles and states, one cannot use the mass-radius observations for specifying their microscopic compositions \textit{directly}. As the most promising sites for the \textit{direct} probe of high-density matter, observations of NS age and surface temperature have been considered for many years (e.g., for a review, see \cite{2004ARA&A..42..169Y,2006NuPhA.777..497P,2021PrPNP.12003879B}). 

Such temperature observations reflect on how isolated NSs cool down mainly due to the loss of neutrinos ($t\lesssim10^5~{\rm yr}$ after their birth), which appear due to various particle scattering inside the NS core. The most widely accepted cooling scenario is the \textit{minimal cooling scenario}, the combination of slow neutrino cooling dominated by baryon bremsstrahlung and modified Urca (MU) process, and enhanced cooling due to baryon superfluidity/superconductivity (here after we simply refer to as SF), so-called pair breaking and formation, which can account for most temperature observation data~\cite{Page2004,2004A&A...423.1063G}, such as the observed cooling rate of Cassiopeia A~\cite{2011PhRvL.106h1101P,2011MNRAS.412L.108S}. But to explain several cold NSs, strong neutrino cooling beyond the minimal cooling scenario is necessary. This statement has been confirmed by recent analysis of XMM--Newton and Chandra data ~\citep{2024NatAs...8.1020M}.


The discovery of cold NSs may imply the possibility of exotic particles, depending on the equation of state (EoS). Within the framework of standard nuclear matter with neutrons, protons, electrons, and muons, the possible fast cooling process is the nucleon direct Urca ($np$DU) process, (anti-)neutrinos emissions through $\beta$ decays. If the density increases, another Urca cooling process could become open due to a higher Fermi level, allowing the appearance of exotic particles such as hyperons, quarks, meson condensation, and so on. In this work, we consider the exotic matter with hyperons and kaon condensation (KC), i.e., the Y+K phase.

Possible existence of KC, i.e., Bose-Einstein condensation of kaons in neutron stars, has been studied extensively from both theoretical and observational aspects~\cite{1986PhLB..175...57K,mtt1993,chlee1996,2000csnp.conf.....G,2020PrPNP.11203770T}. Subsequently, the coexistent phase of KC and hyperons has been elaborated as a realistic form of dense hadronic matter with multi-strangeness.


The possibility of KC in NSs is relevant to the nuclear experiments. For example, the existence of deeply bound kaonic nuclear states has been advocated theoretically based on the strong $\bar K$-nucleon attraction~\cite{Akaishi_2002,Yamazaki_2004}. Recently, formation of basic kaonic clusters, $\bar K N N $, has been reported in the E27 and E15 experiments at J-PARC~\cite{ichikawa_2015,sada_2016,ajimura_2019,yamaga_2020}. 
Multi-$\bar K$ bound nuclear states have also been elucidated theoretically toward future experimental achievement.  
Such objects may be closely related with KC in NSs and may serve to clarify characteristic features of KC and its existence or non-existence in NSs.

To apply the knowledge of KC to NS EoS, an issue came up the softening effect on the EoS, which significantly reduces the maximum mass enough to be incompatible with observed massive NSs~\cite{1994PhRvC..50.3140F,1995PhRvL..75.4567P,1999PhRvC..60b5803G,2000NuPhA.674..553P,2005PhRvC..72c5802M}. Such the softening effect is more highlighted in the case of hyperons (\textit{hyperon puzzle}). Thanks to the recent development of a theoretical framework, however, 2$M_\odot$ NSs can be supported even with the inclusion of KC and hyperons, in the framework of relativistic mean field (RMF) theory. As the phenomenological way to make the EoS stiff in high-density regions, most previous studies have \textit{adjusted} the RMF Lagrangian terms, such as the introduction of density-dependent coupling terms \cite{2014PhRvC..89d5803P,2020PhRvD.102l3007T,2021arXiv210208787B} and mean-field self-interacting $\sigma$ potential \cite{2022PhRvC.105a5807M}, whose physical origins are somewhat unknown. Recently, one of our authors (T.M.) has constructed the stiff Y+K EoS by introducing the \textit{physically-obtained} three-body potential~\cite{2021PhLB..82036587M,2022PTEP.2022i3D03M,2025PhRvC.111d5802M}, which has been supported to exist from the properties of nuclear-matter saturation and light nuclei (for reviews, see \cite{2009PrPNP..62..427L,2021PhR...890....1H}). 


Using the most physically sophisticated Y+K EoS, we investigate the impact of strangeness (hyperons and KC) on neutron-star cooling behavior. Previous studies have investigated the cooling behavior in hyperonic matter~\cite{2009ApJ...691..621T,Negreiros2018,2018MNRAS.475.4347R,2019MNRAS.487.2639R,2021PhRvD.103h3004F,2021PhRvD.103h3004F} and KC matter~\cite{1988PhRvD..37.2042B,2001ApJ...553..382P,2019JKPS...74..547L}, although the calculation of cooling curves with both hyperons and KC has not been done so far (but see the very recent work \cite{2026arXiv260322164W}). In particular, several studies show that proton SF is crucial to see the signature of hyperon DU from cold-NS observations \cite{Negreiros2018,2018MNRAS.475.4347R}. Focusing on proton SF in high-density regions, we explore the possibility of being able to see the signature of hyperons and KC from observed cold isolated NSs.



This paper is structured as follows: In Sect.~\ref{sec:EoS}, we briefly review our Y+K EoS based on \cite{2025PhRvC.111d5802M}. In Sect.~\ref{sec:emissivity}, we describe the properties of neutrino emission processes that occur in Y+K EoS, including baryon SF models. In Sect.~\ref{sec:res}, we present cooling behavior with Y+K EoS, focusing on the role of fast cooling processes and proton SF. Finally, Sect.~\ref{sec:con} is devoted to the conclusion. In Appendix A, the derivation of the neutrino emissivity for the Kaon Urca process is overviewed.

\section{Equation of State with the Y+K phase}
\label{sec:EoS}

\subsection{Overview of formulation}
\label{subsec:formulation}

\subsubsection{
Kaon dynamics in chiral symmetry and baryon-meson interaction in the minimal RMF}
\label{sec:eos}


The Y+K phase is composed of kaon condensates and hyperon-mixed baryonic matter together with  leptons, being kept in beta equilibrium, charge neutrality, and baryon number conservation. In the following, we simply take into account protons, neutrons, $\Lambda$, $\Sigma^-$, and $\Xi^-$ hyperons for baryons and electrons and muons for leptons. 

We base our model for $K$-$B$ and $K$-$K$ interactions upon the effective chiral SU(3)$_{\rm L}$$\times$SU(3)$_{\rm R}$ Lagrangian~\cite{1986PhLB..175...57K,2021PhLB..82036587M,2025PhRvC.111d5802M}. 

In this framework, the nonlinear representation of the charged kaon field $U$ is given by $\xi^2$ 
with
$\xi\equiv \exp[\sqrt{2} i (K^+T_{4+i5} + K^- T_{4-i5})/f]$, where $T_{4\pm i5}$ ($\equiv T_4\pm i T_5$) is the SU(3) generators and $f$ (=93~MeV) is the meson decay constant. The condensed kaon field is assumed to be spatially uniform with spatial momentum ${\bf k}=0$ and represented classically as 
 \begin{equation}
K^\pm =\frac{f}{\sqrt{2}}\theta\exp(\pm i\mu_K t) \ , 
\label{eq:kfield}
\end{equation}
where $\theta$ is the chiral angle, and $\mu_K$ is the $K^-$ chemical potential. 

We adopt the RMF model for two-body $B$-$B$ interaction mediated by the mesons $M$ mediating the baryonic forces: the scalar ($\sigma$, $\sigma^\ast$) mesons and vector ($\omega$, $\rho$, $\phi$) mesons, discarding the {\it nonlinear} self-interacting $\sigma, \omega$, or $\omega-\rho$ meson-coupling potentials. 
We call this model a minimal RMF (MRMF)~\cite{2021PhLB..82036587M,2025PhRvC.111d5802M}. 

 The  Lagrangian density describing the $K$-$B$ and $B$-$M$ interactions is separated into the $s$-wave kaon part ${\cal L}_K$ and the baryon part ${\cal L}_{B,M}$ in the mean-field approximation. 

For ${\cal L}_K$ one reads by the use of Eq.~(\ref{eq:kfield}),
\begin{equation}
{\cal L}_K=f^2\Big\lbrack\frac{1}{2}(\mu_K\sin\theta)^2 - m_K^2(1-\cos\theta)
+2 \mu_K X_0 (1-\cos\theta)\Big\rbrack \ ,
\label{eq:lagk}
\end{equation}
where $m_K$ (=493.68 MeV) is the free kaon mass, and the last term in the bracket stands for the $s$-wave $K$-$B$ vector interaction with $X_0$ being given by 
\begin{eqnarray}
X_0&\equiv&\frac{1}{2f^2}\sum_{b=p,n,\Lambda, \Sigma^-, \Xi^-} Q_V^b n_b \cr
&=& \frac{1}{2f^2}\left(n_p+\frac{1}{2} n_n-\frac{1}{2} n_{\Sigma^-}-n_{\Xi^-} \right)  \ , 
\label{eq:x0}
\end{eqnarray}
where $n_b$ and $Q_V^b$ $\equiv \frac{1}{2}\left(I_3^{(b)}+\frac{3}{2}Y^{(b)}\right)$ are the number density and V-spin charge for baryon species $b$ with $I_3^{(b)}$ and $Y^{(b)}$ being the third component of the isospin and hypercharge, respectively. The form of Eq.~(\ref{eq:x0}) for $X_0$ is specified model-independently within chiral symmetry. 
From Eqs.~(\ref{eq:lagk}) and (\ref{eq:x0}), one can see that the $s$-wave $K$-$B$ vector interaction works attractively for protons and neutrons, while repulsively for $\Sigma^-$ and $\Xi^-$ hyperons, as far as $\mu_K > 0$, and there is no $s$-wave  $K$-$\Lambda$ vector interaction.  

For ${\cal L}_{B,M}$ one reads 
\begin{eqnarray}
{\cal L}_{B,M}&=&\sum_{b=p,n,\Lambda,\Sigma^-,\Xi^-}\overline{\psi}_b \left(i\gamma^\mu D_\mu^{(b)}-M_b^\ast \right) \psi_b \cr
&+&\frac{1}{2}\left(\partial^\mu\sigma\partial_\mu\sigma - m_\sigma^2\sigma^2\right) 
+\frac{1}{2}\left(\partial^\mu\sigma^\ast\partial_\mu\sigma^\ast-m_{\sigma^\ast}^2\sigma^{\ast 2}\right) \cr\cr
&-&\frac{1}{4}\omega^{\mu\nu}\omega_{\mu\nu}+\frac{1}{2}m_\omega^2\omega^\mu\omega_\mu 
-\frac{1}{4}R_a^{\mu\nu} R^a_{\mu\nu}+\frac{1}{2}m_\rho^2 R_a^\mu R^a_\mu \cr\cr
&-& \frac{1}{4}\phi^{\mu\nu}\phi_{\mu\nu}+\frac{1}{2}m_\phi^2\phi^\mu\phi_\mu \ ,
\label{eq:lagbm}
\end{eqnarray}
where the covariant derivative, 
  $\partial_\mu\rightarrow D_\mu^{(b)}\equiv \partial_\mu+i g_{\omega b}\omega_\mu+i g_{\rho b} {\hat I}_3^{\ (b)} R_\mu^3 +ig_{\phi b}\phi_\mu$, is introduced with the vector meson fields for $\omega$, $\rho$, $\phi$ mesons
being denoted as $\omega^\mu$, $R_a^\mu$ with the isospin component $a$, and $\phi^\mu$ ($\sim\bar s\gamma^\mu s$) with ideal mixing, respectively. In the expression of the $D_\mu^{(b)}$, $g_{mb}$ is the vector meson-$B$ coupling constant and ${\hat I}_3^{\ (b)}$ is a sign of the third component of the isospin for baryon $b$.
Throughout this paper, only the time-components of the vector mean fields, $\omega_0$, $R_0$ ($\equiv R_0^{\rm 3}$), $\phi_0$, are considered for description of the ground state and are taken to be uniform. 
The meson masses are set to be $m_\sigma$ = 400 MeV, $m_{\sigma^\ast}$ = 975 MeV, $m_\omega$ = 783 MeV, 
$m_\rho$ = 769 MeV, and $m_\phi$ = 1020 MeV. 
In (\ref{eq:lagbm}), $M_b^\ast$ is the effective baryon mass: 
\begin{equation}
M_b^\ast\equiv M_b -g_{\sigma b}\sigma-g_{\sigma^\ast b}\sigma^\ast -\Sigma_{Kb}(1-\cos\theta)
\ ,
\label{eq:effbm}
\end{equation}
where $M_b$ is the rest mass of baryon $b$, i.e., $M_p$=938.27 MeV, $M_n$=939.57 MeV, $M_\Lambda$=1115.68 MeV, $M_{\Sigma^-}$=1197.45 MeV, and $M_{\Xi^-}$=1321.71 MeV. 
The fourth term on the RHS in Eq.~(\ref{eq:effbm}) represents modification of the free baryon mass $M_b$ through the $s$-wave $K$-$B$ scalar interaction simulated by  the ``kaon-baryon sigma terms'' $\Sigma_{Kb}$ ($b=p, n, \Lambda, \Sigma^-, \Xi^-$), which can be read off from the explicit chiral symmetry breaking terms in the next to leading order $O(m_K^2)$ in chiral expansion with the coefficients $a_1$, $a_2$, and $a_3$ in the effective chiral Lagrangian, 
\begin{subequations}\label{eq:kbsigma}
\begin{eqnarray}
\Sigma_{Kn}&=&-(a_2+2a_3)(m_u+m_s) = \Sigma_{K\Sigma^-} \ ,\label{eq:kbsigma1} \\
\Sigma_{K\Lambda}&=& -\left(\frac{5}{6}a_1+\frac{5}{6}a_2+2a_3\right)(m_u+m_s) \ , \label{eq:kbsigma2} \\
\Sigma_{Kp}&=&-(a_1+a_2+2a_3)(m_u+m_s) = \Sigma_{K\Xi^-} . \label{eq:kbsigma3}
\end{eqnarray}
\end{subequations}
Following a series of our works~\cite{2021PhLB..82036587M,2022PTEP.2022i3D03M,2025PhRvC.111d5802M}, the ``$K$-neutron sigma term'', $\Sigma_{Kn}$, is taken to be 300 and 400~MeV as typical values throughout this paper.  The corresponding depth of the kaon optical potential at the nuclear saturation density $n_0$ (=0.16~fm$^{-3}$), defined by $U_K = \Pi_K(\omega_K; n_{\rm B})/2\omega_K|_{n_{\rm B}=n_0}$ with the self-energy of the $K^-$ meson $\Pi_K(\omega_K; n_{\rm B})$ and the lowest $K^-$ energy $\omega_K$ is $U_K$ = $-$111~MeV and $-$131~MeV, respectively. The main driving forces for the $s$-wave KC are composed of the $KN$ vector interaction ($N = p, n$) [Eq.~(\ref{eq:x0})] and the $KB$ scalar interaction being simulated by $\Sigma_{Kb}$ ($b$=$p, n, \Lambda, \Sigma^-, \Xi^-$) [Eq.~(\ref{eq:kbsigma})], both of which are included in $U_K$.

\subsubsection{Three-baryon forces}
\label{subsubsec:TBR}

 In Eq.~(\ref{eq:lagbm}), there is no extra term with nonlinear self-interacting meson potentials. Instead, many-body baryon interactions, which should be relevant to the stiffness of the EoS in high densities, are introduced by the phenomenological three-baryon forces in the present framework. 

The energy contribution from the three-baryon repulsion is assumed to be qualitatively independent on spin-flavor of baryons, reflecting the confinement mechanisms of quarks at high-density region. Along with this viewpoint, we adopt a specific model for the universal three-baryon repulsion (UTBR) proposed by Tamagaki based on the string-junction model (SJM2) ~\cite{Tamagaki_2008,TakatsukaAIP_2008}.   
We utilize the density-dependent effective two-body potential $U_{\rm SJM}(1,2;n_{\rm B})$ between baryons 1 and 2, by integrating out variables of the third baryon participating the UTBR.
The short-range correlation between baryons is taken into account by way of the correlation function $f_{\rm src}(r)$, 
which is substituted by the result of the reaction matrix calculation in neutron matter by the use of the one-pion exchange Gaussian (OPEG)-A potential~\cite{Tamagaki_2008,TakatsukaAIP_2008}. 
One obtains the approximate form of $U_{\rm SJM}$ as  
\begin{equation}
U_{\rm SJM2}(r; n_{\rm B}) = V_r n_{\rm B}(1+c_r n_{\rm B}/n_0)\exp[-(r/\lambda_r)^2)] \ ,
\label{eq:aUTBR}
\end{equation}
where $V_r$=95 MeV$\cdot$fm$^3$, $c_r$=0.024, and $\lambda_r$=0.86 fm are used corresponding to $\eta_c$ = 0.50 fm for SJM2. The $U_{\rm SJM}$ grows almost linearly with $n_{\rm B}$. 
Finally one obtains the effective two-body potential, 
$\widetilde U_{\rm SJM2}(r;~n_{\rm B})~=f_{\rm src}(r) U_{\rm SJM2}(r; n_{\rm B})$. 

\subsubsection{Three-nucleon attractive force}
\label{subsubsec:TNA}

To simulate the attractive contribution from the three-nucleon attractive force (TNA) to the binding energy for $n_{\rm B}\lesssim n_0$ , we adopt the density-dependent effective two-body potential by Nishizaki, Takatsuka and Hiura~\cite{Nishizaki_1994},\begin{equation}
U_{\rm TNA}(r; n_{\rm B})=V_a n_{\rm B} \exp(-\eta_a n_{\rm B})\exp[-(r/\lambda_a)^2]
(\vec{\bf\tau}_1\cdot\vec{\bf\tau}_2)^2 \ ,
\label{eq:tna}
\end{equation}
where the range parameter $\lambda_a$ is fixed to be 2.0 fm. 
The $U_{\rm TNA}(r; n_{\rm B})$ depends upon not only density but also isospin $\vec \tau_1\cdot\vec \tau_2$ with Pauli matrices $\vec \tau_i$. The parameters $V_a$ and $\eta_a$ are determined together with other parameters to reproduce the saturation properties of symmetric nuclear matter (SNM) for the allowable values of $L$. 

It has been demonstrated that both the UTBR and TNA are necessary in order to reproduce the saturation properties of the SNM in line with the conventional nuclear matter theory, where two-body attraction has a main contribution to the saturation energy and the TNA and three-nucleon repulsion (TNR) have complementary contribution to the energy~\cite{2021PhLB..82036587M,2022PTEP.2022i3D03M,2025PhRvC.111d5802M}. 
 The UTBR has also been shown to be important at high densities as a solution to the hyperon puzzle, making the EOS stiff enough to be consistent with the observations of massive neutron stars. However, in the case of the smaller choice of the slope $L$, there appears a causal limit density, beyond which the sound velocity exceeds the speed of light, but only at a central density of neutron stars near the maximum mass. This is an issue to be resolved in future work. 

There are other baryon interaction models within the RMF which is extended to incorporate the nonlinear self-interacting (NLSI) meson potentials~\cite{Boguta_1977,Serot_1977} or density-dependent (DD) meson-nucleon couplings~\cite{Typel_1999,Char_2014}. 
As for the former (MRMF+NLSI), it should be noted that the many-baryon forces induced by the NLSI terms lead to a quite sizable repulsive contribution for the saturation mechanisms in contrast to the case of the conventional nuclear matter theory. 
In addition, the repulsive energy by the NLSI terms decreases with density at high densities, so that, in the context of the hyperon puzzle, the NLSI terms play a minor role on stiffening of the EOS.  As for the latter (MRMF+DD), there is an ambiguity how the density-dependence of the meson-nucleon couplings is determined microscopically, nevertheless the constructed EOS can become stiff enough to be consistent with the observations of massive neutron stars.

\subsubsection{Energy density expression for the ($Y$+$K$) phase}
\label{subsubsec:energy}

The total energy density ${\cal E}$ is given by
\begin{equation}
{\cal E}={\cal E}_K+{\cal E}_{B,M}+{\cal E}({\rm UTBR})+{\cal E}({\rm TNA})+{\cal E}_{\rm leptons} \ .
\label{eq:total-edensity}
\end{equation} 
From (\ref{eq:lagk}) and (\ref{eq:lagbm}) one obtains
\begin{equation}
{\cal E}_K=\frac{1}{2}(\mu_K f\sin\theta)^2+f^2m_K^2(1-\cos\theta) \ , 
\label{eq:ekfinal}
\end{equation}
\vspace{-0.5cm}~
\begin{eqnarray}
{\cal E}_{B,M}&=&\sum_b \frac{2}{(2\pi)^3}\int_{|{\bf p}|\leq p_F(b)} d^3|{\bf p}|(|{\bf p}|^2+ M_b^{\ast 2})^{1/2} \cr
&+&\frac{1}{2}\left(m_\sigma^2\sigma^2+m_{\sigma^\ast}^2\sigma^{\ast 2}\right) \cr\cr
&+& \frac{1}{2}\left(m_\omega^2\omega_0^2+m_\rho^2 R_0^2+m_\phi^2\phi_0^2\right) \ ,  
\label{eq:ebm}
\end{eqnarray}
where baryons ($b$) are occupied over each Fermi sphere with Fermi momentum $p_F(b)$. 
 
The contribution from the UTBR is written in the Hartree approximation as
\begin{equation}
{\cal E}~({\rm UTBR}) 
=  \frac{\pi^{3/2}}{2}V_r (\widetilde \lambda_r)^3 n_{\rm B}^3\left(1+c_r\frac{n_{\rm B}}{n_0}\right) \ ,  
\label{eq:edUTBRtil}
\end{equation}
where $\displaystyle (\widetilde\lambda_r)^3\equiv \frac{4}{\pi^{1/2}}\int_0^\infty dr r^2f_{\rm src}(r)e^{-(r/\lambda_r)^2}$ (=0.589496 $\cdots$fm$^3$) for SJM2. 

The energy-density contribution from the direct term of the TNA is represented as
\begin{equation}
{\cal E}~({\rm TNA})
=\gamma_a n_{\rm B}^3e^{-\eta_a n_{\rm B}}\lbrace 3-2(1-2Y_p)^2\rbrace 
\label{eq:edTNA}
\end{equation}
with $\displaystyle\gamma_a\equiv(\pi^{3/2}/2) V_a\lambda_a^3$ and $Y_p=n_p/n_{\rm B}$ the proton-mixing ratio. 

For leptons, the energy density contribution from the ultra-relativistic electrons is given as
\begin{equation}
{\cal E}_e\simeq\mu_e^4/(4\pi^2) \ . 
\label{eq:ee}
\end{equation}
In case $\mu_\mu > m_\mu$ with $\mu_\mu$ being the muon chemical potential and $m_\mu$ the muon mass (=105.66 MeV), there appear muons in the ground state and participate in charge neutrality and $\beta$ equilibrium conditions. With the muon Fermi momentum $p_F(\mu^-)$, the energy density for muons are given as
\begin{eqnarray}
\hspace{-0.8cm}{\cal E}_\mu&=&\frac{2}{(2\pi)^3}\int_{|{\bf p}|\leq p_F(\mu^-)} d^3|{\bf p}|(|{\bf p}|^2+m_\mu^2)^{1/2} \cr
&=&\frac{m_\mu^4}{8\pi^2}\Big\lbrack r(1+2r^2)\sqrt{r^2+1} 
- \log(r+\sqrt{r^2+1}) \Big\rbrack 
\label{eq:emuon}
\end{eqnarray}
with $r\equiv p_F(\mu^-)/m_\mu$.  

\subsubsection{Ground-state conditions}
\label{subsubsec:grcond}

The ground state energy for the Y+K phase is obtained under the charge neutrality, baryon number, and $\beta$-equilibrium conditions. The charge neutrality condition is written as  
\begin{equation}
n_Q=n_p-n_{\Sigma^-}-n_{\Xi^-}-n_{K^-}-n_e-n_\mu=0 \ , 
\label{eq:charge}
\end{equation}
where $n_Q$ denotes the total negative charge density, $n_{K^-}$ is the number density of KC and is given from kaon part of the Lagrangian density (\ref{eq:lagk}) as
\begin{eqnarray}
n_{K^-}&=&-iK^-(\partial{\cal L}_K/\partial\dot{K^-})+iK^+(\partial{\cal L}_K/\partial\dot{K^+}) \cr
&=&\mu_K f^2\sin^2\theta+2f^2X_0(1-\cos\theta) \ . 
\label{eq:rhokc}
\end{eqnarray}
In Eq.~(\ref{eq:charge}), $n_e$ is the electron number density and is related to the electron chemical potential $\mu_e$ as $n_e=\mu_e^3/(3\pi^2)$ in the ultra-relativistic limit. $n_\mu$ is the muon number density and is given by $n_\mu=[p_F(\mu^-)]^3/(3\pi^2)$. 

The baryon number conservation is given by
\begin{equation}
n_p +n_n + n_\Lambda+ n_{\Sigma^-}+ n_{\Xi^-}= n_{\rm B} \ .
\label{eq:bn}
\end{equation}
In addition, the following chemical equilibrium conditions for weak processes are imposed:
 $n\rightleftharpoons p+K^-$, $n\rightleftharpoons p+e^-$, $n + e^-\rightleftharpoons \Sigma^-$, $\Lambda + e^-\rightleftharpoons \Xi^-$, $n\rightleftharpoons \Lambda$, and those involved in muons in place of $e^-$ if muons are present. 
 These conditions are followed by the relations between the chemical potentials 
\begin{eqnarray}
\mu=\mu_K&=&\mu_e=\mu_\mu=\mu_n-\mu_p \ , \cr
\mu_\Lambda&=&\mu_n , \cr
 \mu_{\Sigma^-}&=&\mu_{\Xi^-}=\mu_n+\mu_e \ ,
\label{eq:chem}
\end{eqnarray}
where $\mu$ and $\mu_i$ (=$\partial{\cal E}/\partial n_i$) ($i$= $p$, $n$, $\Lambda$, $\Sigma^-$, $\Xi^-$, $K^-$, $e^-$, $\mu^-$) are the charge chemical potential and the chemical potential for each particle species ($i$), respectively, at a given baryon number density $n_{\rm B}$. 

\subsubsection{Determination of parameters}
\label{subsec:para}

{\it Meson-nucleon coupling constants from saturation properties in symmetric nuclear matter}

The empirical nuclear saturation density $n_0$ is set to 0.16~fm$^{-3}$. Then 
the meson-nucleon coupling constants, $g_{\sigma N}$, $g_{\omega N}$, $g_{\rho N}$, the meson mean-fields at $n_0$, $\langle\sigma\rangle_0$, $\langle\omega_0\rangle_0$, and the parameters $\gamma_a$, $\eta_a$ in TNA are determined from the saturation properties of SNM, i~.e., the nuclear binding energy per unit of nucleon at $n_0$, $B_0$ = $-$16.3~MeV, the equations of motion for the $\sigma$ and $\omega_0$ fields at $n_0$,  together with the empirical values of the incompressibility $K_0$=240 MeV~\cite{Garg_2018}, the symmetry energy $S_0$ (=31.5 MeV) at $n_0$~\cite{Li_2013}, and the slope $L$ of the symmetry energy, which is defined as  
$L$$\equiv 3 n_0\left(d S(n_{\rm B}) / d n_B\right)_{n_B=n_0}$ with $S(n_{\rm B})$ being the symmetry energy as a function of baryon density. 
The empirical value of the $L$ has a large uncertainty, ranging from 30 MeV to 90 MeV~\cite{Lattimer_2016,Oertel_2017}. In Ref.~\cite{2025PhRvC.111d5802M}, $L$ is selected within the range (60$-$70)~MeV, for which the resulting EOS based on our interaction model leads to have the Y+K core in massive compact stars, keeping the EOS stiff enough to be consistent with recent observations of massive compact stars. Further, within this allowable range for $L$, the EOS with the Y+K phase does not become much different quantitatively. Hence, we take the typical value $L$ = 65 MeV throughout this paper.

\subsubsection{Meson-hyperon coupling constants}
\label{subsubsubsec:MYcoupling}

For the description of hyperon-mixed matter, we set the values of the meson-hyperon coupling constants to obtain the hyperon-nucleon and hyperon-hyperon interactions in the MRMF. 

The vector meson couplings for hyperons ($Y$) are obtained from the vector-nucleon couplings $g_{\omega N}$, $g_{\rho N}$, $g_{\phi N}$ through the SU(6) symmetry relations~\cite{sdg_1994} :  \begin{subequations}
\begin{eqnarray}\label{eq:gmY}
g_{\omega\Lambda}&=&g_{\omega\Sigma^-}=2g_{\omega \Xi^-}=(2/3) g_{\omega N} \ , 
\label{eq:gmY1} \\
g_{\rho \Lambda}&=& 0 \ , g_{\rho\Sigma^-}=2g_{\rho\Xi^-}=2g_{\rho N} \ , \label{eq:gmY2} \\ 
 g_{\phi\Lambda}&=& g_{\phi\Sigma^-}=(1/2) g_{\phi\Xi^-}=-(\sqrt{2}/3) g_{\omega N} \ . 
\end{eqnarray}
\end{subequations}
The scalar ($\sigma$, $\sigma^\ast$) meson-hyperon couplings are determined from the phenomenological analyses of recent hypernuclear experiments. 

From recent theoretical and experimental analyses on hypernuclear experiments~\cite{Gal_2016}, the values of the $V_Y^N$ are set to be $V_\Lambda^N=-27$ MeV, $V_{\Sigma^-}^N$ = 23.5 MeV, and $V_{\Xi^-}^N$ = $-14$ MeV, and one obtains $g_{\sigma\Lambda}$, $g_{\sigma\Sigma^-}$, and $g_{\sigma\Xi^-}$. These coupling constants are obtained for each case of $L$.
For the details of estimating the $g_{\sigma Y}$ and $g_{\sigma^\ast Y}$, 
see Ref.~\cite{2022PTEP.2022i3D03M}. 

\subsection{Properties of compact stars with the Y+K phase}
\label{subsec:results-EoS}

In Fig.~\ref{fig:M-rhoc}, we show the gravitational mass $M/M_\odot$ as functions of mass density at the center of the star, $\rho_{\rm center}$ ($\equiv {\cal E}_{\rm center}/c^2$ with $ {\cal E}_{\rm center}$ being the energy density at the center and $c$ the speed of light).
Different EoS models are shown for $L$ = 65~MeV; The branches including KC in the core are denoted as the bold solid lines (thin solid lines) for $\Sigma_{Kn}$ = 300 MeV ($\Sigma_{Kn}$ = 400 MeV).  For comparison, the branch with pure hyperonic matter (abbreviated to PYM), where KC is switched off by setting $\theta=0$, is shown by the green dashed line. 

The filled triangle [$\blacktriangle$] stands for the branch point where the $\Lambda$ hyperons appear from nuclear matter in the center of the star. The branch point at which KC appears in the center of the star is indicated by the filled circle [$\bullet$] (open circle [$\circ$]) in the case of $\Sigma_{Kn}$ = 300 MeV (400 MeV). Also the branch point at which $\Xi^-$ hyperons appear in the center of the star is indicated by the filled inverted triangle [$\blacktriangledown$] (open inverted triangle [$\triangledown$]) in the case of $\Sigma_{Kn}$ = 300 MeV ($\Sigma_{Kn}$ = 400 MeV). 
 The cross point [$\times$] denote the causal limit beyond which the sound velocity exceeds the speed of light. 
The density at which the center of the star reaches the onset of the $np$DU process is denoted by the asterisk. In Table~\ref{tab:onset}~(a), the numerical values of onset densities for the $np$DU process, the $\Lambda$, $\Xi^-$ hyperons, and KC are listed in the unit of nuclear saturation mass density $\rho_0$ for each case of PYM, $\Sigma_{Kn}$ = 300~MeV, and $\Sigma_{Kn}$ = 400~MeV. In Table~\ref{tab:onset}~(b), mass densities at the center of compact stars, $\rho_{\rm center}$, in the unit of $\rho_0$ with some typical gravitational masses, 1.4~$M_\odot$, 1.8~$M_\odot$, and the maximum mass $M_{max}$ are listed. It is to be noted that, as discussed in Ref.~\cite{2021PhLB..82036587M,2025PhRvC.111d5802M}, our EOS with the Y+K phase is consistent with observational results on the mass and radius for several neutron stars by the Neutron star Interior Composition Explorer (NICER).


\section{Neutrino emissivities}
\label{sec:emissivity}

In light NSs, the cooling curves obey the minimal cooling scenario, which consists of slow cooling processes (mainly MU process and baryon bremsstrahlung) and pair breaking and formation (PBF) due to the transition to baryon SF phase \cite{Page2004,2004A&A...423.1063G}. These neutrino emissivities can be roughly expressed as
\begin{eqnarray}
    \epsilon_{\nu,{\rm slow}}^0&\approx&10^{19-21}(T/10^9~{\rm K)}^6~{\rm erg~cm^{-3}~s^{-1}}~, \nonumber \\
    \epsilon_{\nu,{\rm PBF}}&\approx&10^{22}T^7F(T/T_c)~{\rm erg~cm^{-3}~s^{-1}}~, \label{eq:minimal}
\end{eqnarray}
respectively. Here, $F$ denotes the efficiency of PBF as a function of $T/T_{c,b}$ where $T_{c,b}$ denotes
SF transition temperature of corresponding baryons \cite{Page2004}. The neutrino emissivities with subscript "0" ($\epsilon^0$) denote those in non-SF NSs, and could be significantly reduced if SF baryons exist. The minimal cooling scenario can account for most temperature observations of isolated NSs, except for several cold stars such as PSR J0205$+$6449, PSR B2334$+$61, and CXOU J8052$-$4617 (or Vela Jr.)~\cite{2024NatAs...8.1020M}: This issue is solved by considering fast cooling processes resulting in drastic surface-temperature drop around $t\sim100~{\rm yr}$, for which are there many candidates as a representative of nucleon direct Urca (DU) process. Unlike the slow cooling processes, there are threshold conditions for fast cooling processes to occur, because of fewer number of particles involved, which makes it harder for the kinematic condition to meet, at least for low-mass stars. In high-mass NSs, other baryons or mesons except neutrons are likely to appear, which generally allows the corresponding DU reactions. 

Explicitly important physical ingredients to describe the DU reactions are particle fractions and the effective mass ratio of relevant particles, which are presented in Figs. \ref{fig:yi} and \ref{fig:meff}, respectively. As shown in our previous work~\cite{2025PhRvC.111d5802M}, $K^-$ appears with typically $Y_{K^-}\sim\mathcal{O}(0.1)$ in high-density regions, where $\Xi^-$ does not appear so much (at most $Y_{\Xi^-}<$0.08). These features are reflected in the density dependence of the effective mass: While $K^-$ mass significantly decreases, $\Xi^-$ mass does not. We can also see that the fraction and effective mass of other particles, in particular for $\Lambda$ are not sensitive to $\Sigma_{Kn}$ values, because of no explicit particle transfer between $\Lambda$ hyperons and $K^-$. From these, one can expect rapid cooling associated with $\Lambda$ hyperons and KC. 

Below, focusing on hyperon-mixed KC matter, we describe the properties of DU reactions.


\subsection{Nucleon-Induced Urca process}
\label{sec:NU-DU}

The nucleon DU process is the only fast cooling process that is not involved with exotic matter, since it is the neutrino emission due to the beta decay and inverse beta decay \cite{1981PhLB..106..255B}:
\begin{subequations}\label{eq:DUNuc}
\begin{align}
n &\rightarrow p + l^- + \bar{\nu}_l, \\
p + l^- &\rightarrow n + \nu_l~,
\end{align}
\end{subequations}
where $l = e^-, \mu^-$ denotes leptons. The corresponding emissivity in the absence of baryon pairing is \cite{1991PhRvL..66.2701L} \footnote{We note that, if the KC appears, the expression of the emissivity of $np$DU and $\Lambda p$DU are slightly changed because of
\begin{itemize}
    \item appearance of neutron-to-proton mass difference due to the contribution of the last term in Eq. (5), which is at most 40\% (see Fig. \ref{fig:meff}).
    \item cooling suppression due to reduction of the matrix elements by a factor of $\cos^2(\theta/2)$ for $np$DU ~\cite{1994NuPhA.571..758F} and $\cos^4(\theta/2)$ for $\Lambda p$DU, but not significant \cite{1994PhRvC..50.3140F}.   
\end{itemize}
Formulae of their emissivities, i.e., Eqs. (\ref{eq:$np$DU}) and (\ref{eq:lpdu}), do not consider the above for simplicity.}
\begin{eqnarray}
    \epsilon_{\nu,np{\rm DU}}^0 &=& 4.0\times10^{27}~{\rm erg~cm^{-3}~s^{-1}} \left(\frac{{M}_N^*}{M_N}\right)^2T_9^6 \nonumber \\
  &\times& \left[\left(\frac{n_BY_e}{n_0}\right)^{1/3}\Theta_{npe}+\left(\frac{n_BY_\mu}{n_0}\right)^{1/3}\Theta_{np\mu}\right], \label{eq:$np$DU}
\end{eqnarray}
where $T_9\equiv T/(10^9~{\rm K})$, $Y_i\equiv n_i/n_B$, $M_N$ (=939.57 MeV) the nucleon rest mass, and $M_N^*/M_N$ the effective mass ratio of neutrons and protons. The nucleon DU is stronger by at least five orders of magnitude than the minimal cooling, as we see the coefficient of each emissivity. In Eq.~(\ref{eq:$np$DU}), 
$\Theta_{ijk}\equiv\Theta(|\bm{p}_F(j)+\bm{p}_F(k)|-p_F(i))\cdot\Theta(p_F(i)-|\bm{p}_F(j)-\bm{p}_F(k)|)$, where $\Theta(x)$ is the step function on the conservation of Fermi momenta of $i,j,k$ particles. $\Theta_{ijk}$ determines the threshold NS mass on the onset of the corresponding reaction. This condition on the nucleon DU process, i.e., $p_F(n)<|\bm{p}_F(p)+\bm{p}_F(l)|$, finally reads
\begin{align}
 Y_p>\left[1+\left(1+x_l^{1/3}\right)^3\right]^{-1}
\left(1-Y_\Lambda-Y_{\Sigma^-}-Y_{\Xi^-}\right)~,\label{eq:ypthreal}
\end{align}
where $x_l\equiv Y_l/(Y_e+Y_\mu+Y_{\Sigma^-}+Y_{\Xi^-}+Y_{K^-})$, and $Y_{\Sigma^-}=0$ in the Y+K EOS. Even in the absence of muons, the expression is in a similar form\footnote{Generally, the critical $Y_p$ value increases as is the case of $npe\mu$ matter \cite{2017IJMPE..2650015L,2019PTEP.2019k3E01D,2024PhRvC.110a5805L}.}. As shown in Fig. \ref{fig:M-rhoc}, the threshold mass for the DU process with electrons in Y+K EoS, i.e., the condition to satisfy Eq. (\ref{eq:ypthreal}), is $M_{\rm DU}=1.3~M_\odot$.


Furthermore, if one assumes purely $\Lambda$-hyperon matter, Eq. (\ref{eq:ypthreal}) reduces to
\begin{align}
  Y_p>\frac{1}{9}\left(1-Y_\Lambda\right)~,\label{eq:ypthhyp}
\end{align}
This factor "1/9" is a well-known threshold of proton fraction with $npe$ matter. Thus, it is easy to see that the inclusion of $\Lambda$ hyperons makes the threshold of proton fraction lower. Also, by comparing Eq. (\ref{eq:ypthreal}) and (\ref{eq:ypthhyp}) without $\Xi^-$ hyperons, one can see that KC increases the threshold of proton fraction higher compared to that with purely $\Lambda$-hyperon matter, due to the decrease of $x_l$. This property could become important not in our Y+K EoS but in very small symmetry-energy EoSs, because this allows the DU process to operate only in high-density regions where hyperons appear to a certain extent.
 




\subsection{Lambda ($\Lambda$) Induced URCA process}
\label{sec:LU-DU}
At supranuclear densities, the appearance of exotic degrees of freedom such as the $\Lambda$ hyperon opens additional weak-interaction channels for neutrino and antineutrino production. Being electrically neutral, $\Lambda$ hyperons typically constitute the first hyperonic species to emerge in neutron-star matter because they are the lightest. Once present, $\Lambda$ hyperons participate in beta-type weak processes that closely resemble the nucleon-induced Direct Urca (DU) reactions discussed in Sec.~\ref{sec:NU-DU}. The corresponding kinematic conditions are likewise analogous to those governing the nucleonic DU process.

The dominant $\Lambda$-induced DU reactions can be written as
\begin{subequations}\label{eq:DULambda}
\begin{align}
\Lambda &\rightarrow p + l^- + \bar{\nu}_l, \\
p + l^- &\rightarrow \Lambda + \nu_l~.
\end{align}
\end{subequations}
The corresponding emissivity in the absence of baryon pairing is \cite{1992ApJ...390L..77P}
\begin{eqnarray}
    \epsilon_{\nu,\Lambda p{\rm DU}}^0 &=& 1.6\times10^{26}~{\rm erg~cm^{-3}~s^{-1}}\left(\frac{{M}_N^*}{M_N}\right)\left(\frac{M_\Lambda^*}{M_\Lambda}\right)T_9^6 \nonumber \\
    &\times& \left[\left(\frac{n_BY_e}{n_0}\right)^{1/3}\Theta_{\Lambda pe}+\left(\frac{n_BY_\mu}{n_0}\right)^{1/3}\Theta_{\Lambda p\mu}\right], \label{eq:lpdu}
\end{eqnarray}
where $M_\Lambda^*/M_\Lambda$ is the effective mass ratio of $\Lambda$ hyperons. As with the nucleon DU, $\Lambda$-hyperon DU is also much stronger than the minimal cooling.

The kinematic condition on $\Theta_{\Lambda pl}$ is primary $p_F(\Lambda)<p_F(p)+p_F(l)$\footnote{One may also consider $p_F(p) < p_F(\Lambda) + p_F(l)$, which finally reads
\begin{eqnarray}
 Y_p < Y_{\Lambda}\left(1+\left(\frac{Y_l}{Y_{\Lambda}}\right)^{1/3}\right)^3~,\label{eq:lpduno}
\end{eqnarray}
which is however very lax compared to Eq.~(\ref{eq:lpduthreal}) because the right hand is near unity once $Y_{\Lambda}>0$.
}, which finally reads
\begin{eqnarray}
 Y_p > Y_{\Lambda}\left(1-\left(\frac{Y_l}{Y_{\Lambda}}\right)^{1/3}\right)^3~.\label{eq:lpduthreal}
\end{eqnarray}
In the absence of muons, $\Xi^-$ hyperons and $K^-$ mesons, i.e., $np\Lambda e$ matter, this reduces to 
\begin{eqnarray}
Y_{p}>\frac{1}{8}Y_{\Lambda}~.
\end{eqnarray}
This condition easily meets with few hyperons at most $\sim$0.03 \cite{1992ApJ...390L..77P} (see Fig.~\ref{fig:yi}). Hence, the threshold mass for the hyperon DU process is almost the same as that where $\Lambda$ hyperons appear, unless the symmetry energy is quite low. In Y+K EoS, threshold mass for the $\Lambda$-hyperon DU process is $M_{Y}=1.5~M_\odot$.




\subsection{Xi ($\Xi$) Induced URCA process}

At even higher baryon densities, the appearance of negatively charged cascade hyperons opens additional channels for fast neutrino emission. In hyperon--kaon (Y+K) EoS, instead of $\Sigma^-$, heavier $\Xi^-$ hyperons typically emerge at densities exceeding those required for $\Lambda$ hyperons because of stronger repulsion in $\Sigma^-$--nucleon interaction compared to $\Xi^-$--nucleon (see also Section II-A7). 



The dominant $\Xi^-$-induced DU processes can be expressed as
\begin{subequations}\label{eq:DUXi}
\begin{align}
\Xi^- &\rightarrow \Lambda + l^- + \bar{\nu}_l, \\
\Lambda + l^- &\rightarrow \Xi^- + \nu_l .
\end{align}
\end{subequations}
In principle, these reactions provide an efficient neutrino cooling channel due to their direct nature and the absence of an additional spectator baryon. However, the activation of the $\Xi^-$-induced DU process is strongly constrained by kinematic conditions. This is because $\Lambda$ hyperons are much more than $\Xi^-$ and electrons ($Y_\Lambda\gg Y_{\Xi^-}\sim Y_e\sim\mathcal{O}(0.01)$; see also Fig.~\ref{fig:yi}), showing that the Fermi momentum of $\Lambda$ hyperons significantly exceeds that of electrons and $\Xi^-$ hyperons ($p_F(\Lambda) \gg p_F(\Xi^-)\sim p_F(e)$). As a result, despite the presence of $\Xi^-$ hyperons, the corresponding DU channel is often suppressed or entirely forbidden.

When allowed, the neutrino emissivity of the $\Xi^-$-induced DU process follows the characteristic $T^6$ temperature dependence where the strength is about one-hundredth as strong as the nucleon DU~\cite{1992ApJ...390L..77P}. In that case, it is one of the candidates as the site of rapid cooling, although it is prohibited with Y+K EoS.




\subsection{Kaon-induced Urca process}
\label{subsubsec:KU}

Kaon condensates affect not only the structure of compact stars but also their thermal evolution through rapid cooling via neutrino and anti-neutrino emissions. The extra weak process occurring in the presence of KC is the kaon-induced Urca (KU) process associated with baryon $B$ ($B = p,n, \Lambda,\Xi^-$) (abbreviated to $BB$KU), 
\begin{subequations}\label{eq:KU}
\begin{eqnarray}
&& B + \langle K^-\rangle\rightarrow B + l +\bar\nu_l \ , \label{eq:KUa} \\
&& B + l \rightarrow B + \langle K^- \rangle + \nu_l   \label{eq:KUb}
 \end{eqnarray}
 \end{subequations}
($l = e^-,\mu^-$), where $\langle K^- \rangle$ stands for the classical $K^-$ field which is crucial to supply the system with energy $\mu_K$ and to make the reaction kinematically possible~\cite{Tatsumi_1988,Brown_1988}. 

It has been shown that both weak interaction relevant to cooling processes with KC and strong interaction relevant to the EoS of kaon-condensed phase can be considered in a unified manner within current algebra and partial conservation of axial current (PCAC)~\cite{Tatsumi_1988}. In the following, we only show the result of the emissivity for $NN$KU process.
In Appendix A, the derivation of the neutrino emissivity for the $NN$KU process is outlined in line with chiral symmetry framework with the nonlinear representation of the kaon field. Some hyperon-associated KU processes are also briefly mentioned.

The total emissivity for the KU process including both proton and neutron associated processes is given by
\begin{widetext}
\begin{subequations}\label{eq:NNKU}
\begin{eqnarray}
\epsilon^0_{\nu,pp{\rm KU}}&\equiv&\epsilon^0_{\nu,pp{\rm KU}}~(a)+\epsilon^0_{\nu,pp{\rm KU}}~(b)\simeq (2.6\times 10^{25})\Bigg(\frac{ M_p^\ast}{M_N}\Bigg)^2\frac{\mu_e}{m_\pi}\sin^2\theta T_9^6 \ \  ( {\rm erg}\cdot {\rm cm}^{-3}\cdot{\rm s}^{-1} ) \ , \label{eq:ppKU} \\
\epsilon^0_{\nu,nn{\rm KU}}&\equiv&\epsilon^0_{\nu,nn{\rm KU}}~(a)+\epsilon^0_{\nu,nn{\rm KU}}~(b)\simeq(5.8\times 10^{24})\Bigg(\frac{ M_n^\ast}{M_N}\Bigg)^2\frac{\mu_e}{m_\pi}\sin^2\theta T_9^6 \ \  ( {\rm erg}\cdot {\rm cm}^{-3}\cdot{\rm s}^{-1} ) \ .
\label{eq:nnKU}
\end{eqnarray}
\end{subequations}
\end{widetext}
Here, the charge chemical potential $\mu$ (= $\mu_e=\mu_K$) is scaled by the pion rest mass  $m_\pi$ ($\simeq$140 MeV).
In Eqs.~(\ref{eq:ppKU}) and (\ref{eq:nnKU}), the emissivities for anti-neutrino-emission process (a) and neutrino-emission process (b) are set to be equal since the system is considered to be in beta equilibrium. The numerical value of the emissivity for the process (a) of the $nn$KU (\ref{eq:nnKU}) should be compared with that in ref.~\cite{1995PhRvD..52.3739T,2019JKPS...74..547L}.

In our calculation, the $pp$KU is shown to be stronger by a factor of 4 compared to $nn$KU. As the KU is thought to be insufficient to explain some cold quiescent NSs \cite{2004ARA&A..42..169Y}, this update on $pp$KU may become a key for solving this issue.


\subsection{Suppression of DU/KU cooling by superfluidity}

When NS cools with $T<T_{c,b}$, the corresponding baryons become SF states, which significantly suppresses fast neutrino cooling processes as
\begin{eqnarray}
\epsilon_{\nu,np{\rm DU}} &=& R_{np}(T/T_{c,n},T/T_{c,p})\epsilon_{\nu,np{\rm DU}}^0 \\
    \epsilon_{\nu,\Lambda p{\rm DU}}  &=& R_{\Lambda p}(T/T_{c,\Lambda},T/T_{c,p}) \epsilon_{\nu,\Lambda p{\rm DU}}^0 \\
    \epsilon_{\nu,bb{\rm KU}} &=& R_b(T/T_{c,b})\epsilon_{\nu,bb{\rm KU}}^0~,
\end{eqnarray}
where $R_b$ denotes the reduction factor of the baryon SF, and $R_{b_1b_2}\approx \mathrm{min}(R_{b_1},R_{b_2})$~(Section 4.3 in \cite{2001PhR...354....1Y}). Here, SF neutrons are assumed to become $^3{\rm P}_2$ state while other SF baryons $^1{\rm S}_0$ state \cite{1970PThPh..44..905T,2013arXiv1302.6626P}. Then, the reduction factor for $T\ll T_{c,b}$ would become $\exp(-aT_{c,b}/T)$ where $a=8.40$ for neutrons while $a=1.76$ for other baryons~\cite{1998PhR...292....1T}. The higher $T_{c,b}$ is, the suppression effects on cooling curves work in earlier time, which could be before thermal relaxation timescale ($t\sim100~{\rm yr}$). In this case, because rapid cooling can \textit{vanish} in cooling curves, some observed cold NSs can be explained even with the presence of fast cooling processes.


\begin{figure}[t]
    \centering
    \includegraphics[width=0.9\linewidth]{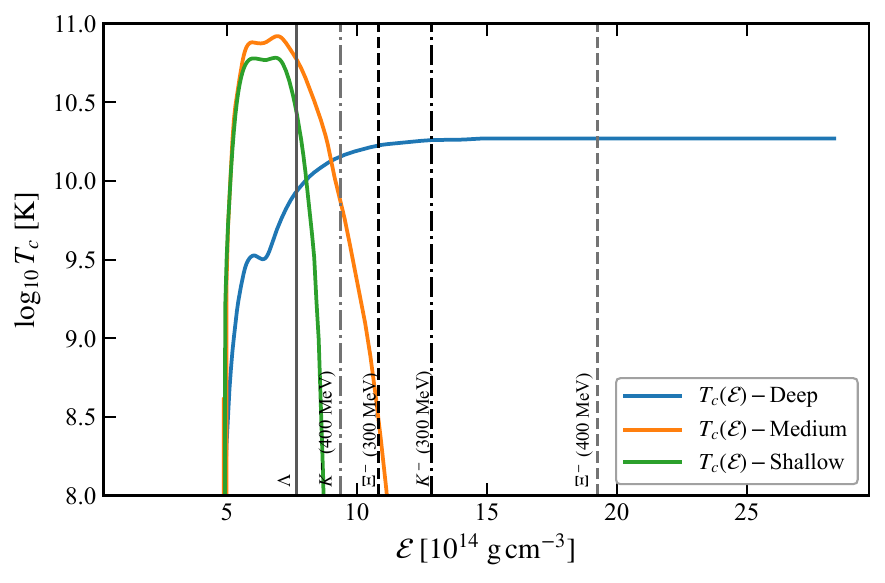}
    \caption{
    $^1{\rm S}_0$ proton SF models utilized in this study. Onset densities of hyperons and KC are also shown.
    }
    \label{fig:tc_p}
\end{figure}

\par The density dependence of $T_{c,b}$ remains highly uncertain owing to uncertainties in the nuclear pairing interaction. For the $^1{\rm S}_0$ and $^3{\rm P}_2$ neutron pairing gaps, we adopt the T72 model \cite{Takatsuka1972} and the ``a'' model of \citet{Page2004}, respectively, and keep these fixed throughout this work. For simplicity, we do not include hyperon superfluidity. Our primary objective is to isolate the role of $^1{\rm S}_0$ proton superconductivity, as it suppresses both the $np$DU and $\Lambda p$DU processes, thereby significantly modifying the fast-cooling scenario. In microscopic many-body calculations, the proton pairing gap is generally expected to decrease with increasing density because repulsive three-body forces become more important and the proton effective mass decreases, reducing the density of states near the Fermi surface. Consequently, whether proton superconductivity survives into the highest-density regions of neutron-star cores remains an open question.

The density dependence of the proton $^1{\rm S}_0$ pairing gap is one of the major theoretical uncertainties in neutron-star cooling calculations. Existing microscopic many-body calculations predict substantially different critical temperatures and density extensions owing to uncertainties in the nuclear interaction, three-body forces, effective masses, and medium polarization effects. In particular, many calculations predict that the proton pairing gap decreases and eventually disappears at supranuclear densities, although the exact density dependence remains uncertain. Consequently, rather than adopting a specific microscopic pairing model, we follow the phenomenological approach of \citet{Negreiros2018}, who introduced representative proton superconductivity models to investigate the sensitivity of neutron-star cooling to the density extent of proton pairing.

Following Ref.~\cite{Negreiros2018}, we employ the same three phenomenological $^1{\rm S}_0$ proton superconductivity models, referred to as the shallow, medium, and deep models, as shown in Fig.~\ref{fig:tc_p}. In the shallow model, $T_{c,p}$ is restricted to relatively low densities of $(2$--$3)\,n_0$. In the medium model, proton superconductivity extends up to $\sim4\,n_0$. In the deep model, it extends beyond $5\,n_0$, while maintaining a lower peak value of $T_{c,p}$ than in the shallow and medium models. The principal difference between these models therefore lies in the high-density region. Whereas many microscopic calculations predict that the $^1S_0$ proton pairing gap decreases and eventually disappears with increasing density owing to repulsive three-body interactions and the reduction of the proton effective mass, the deep model instead assumes that proton superconductivity persists into the high-density core. This model should therefore be regarded as a phenomenological upper-limit scenario, introduced to explore the consequences of extended proton superconductivity on neutron-star cooling, rather than as a definitive prediction of current microscopic many-body calculations.




\begin{figure}
    \centering
    \includegraphics[width=0.9\linewidth]{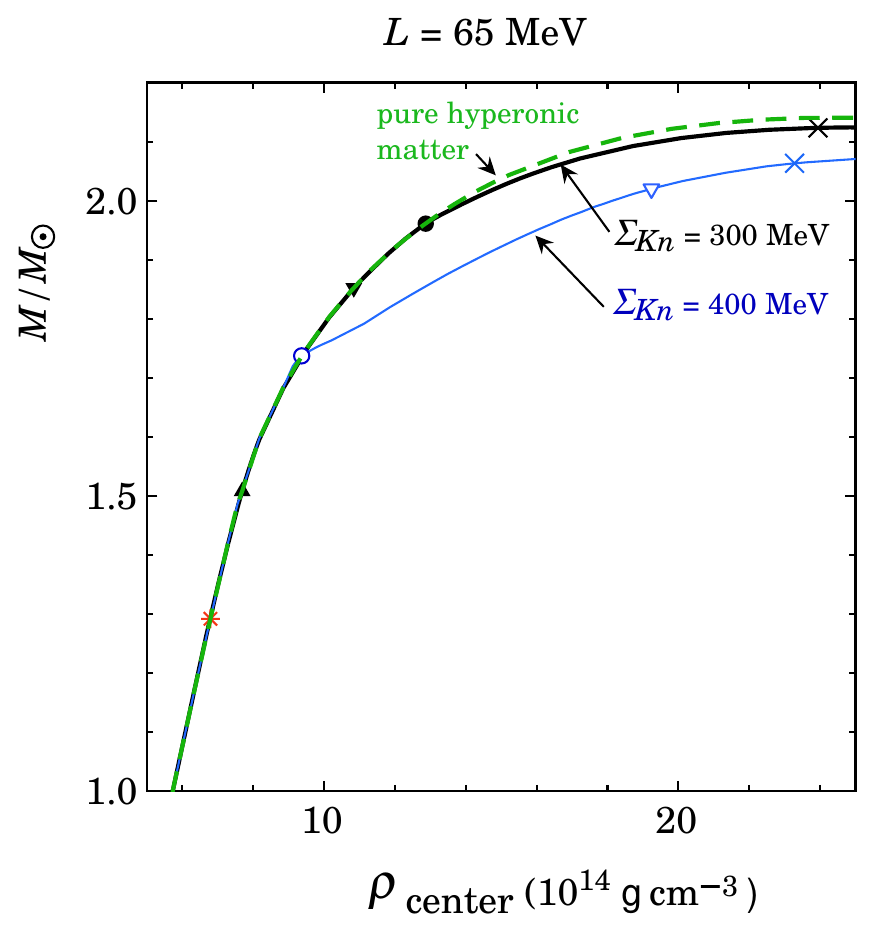}
    \caption{The gravitational mass in the unit of the solar mass, $M/M_\odot$, as functions of density at the center of the star, $\rho_{\rm center}$ ($\equiv {\cal E}_{\rm center}/c^2$). Different EoS models for $L$ = 65~MeV are shown; The branches including KC in the core are denoted as the black bold solid lines (blue thin solid lines) for $\Sigma_{Kn}$ = 300 MeV ($\Sigma_{Kn}$ = 400 MeV).  For comparison, the branch with pure hyperon-mixed matter, where KC is switched off by setting $\theta=0$, is shown by the green dashed line. The density, at which the center of the star reaches the onset of the direct Urca process, is denoted by the red asterisk. See the text for details. \\
    }
    \label{fig:M-rhoc}
\end{figure}
    
\begin{figure*}[t]
    \centering
\includegraphics[width=0.8\linewidth]{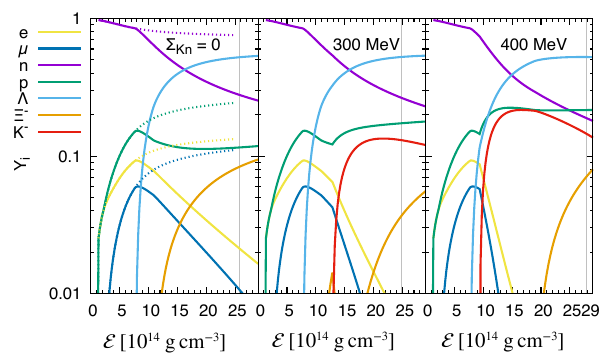}
    \vspace*{-0.5cm}
    \caption{Density dependence of particle fraction with different $\Sigma_{Kn}$ values. Grey lines indicate the central density at the maximum mass. For $\Sigma_{Kn} = 0$, we also show the case of pure nucleonic matter as dotted lines.\\
    }
    \label{fig:yi}
    \centering
\includegraphics[width=0.8\linewidth]{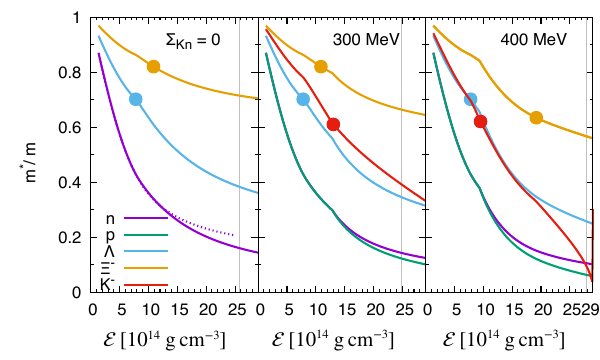}
\vspace*{-0.5cm}
    \caption{Same as Fig. \ref{fig:yi} but for effective masses of each baryon and $K^-$. The circle indicates the onset density of the corresponding particle.}
    \label{fig:meff}
\end{figure*}

\begin{table*}[!]
\caption{(a) The onset densities in the MRMF+ UTBR+TNA model for $\Sigma_{Kn}$ = 300~MeV, and 400~MeV for slope $L$=65~MeV. For comparison, the case of the pure hyperonic matter (PYM), for which the chiral angle $\theta$ for KC is set to zero, is symbolically denoted as $\Sigma_{Kn}$ = 0 for the sake of brevity. 
$\rho^c(\rm DU)/\rho_0$ is the onset density at which the direct Urca process (DU) occurs, 
$\rho^c(\Lambda)$ the onset density of $\Lambda$ hyperons in the normal neutron-star matter, $\rho^c(\Xi^-)$ the one of $\Xi^-$ hyperons in the $\Lambda$-mixed matter, and $\rho^c(K^-)$ is the one of KC in the hyperon ($\Lambda$ and/or $\Xi^-$)-mixed matter. All the densities are denoted in the unit of mass density at nuclear matter saturation $\rho_0$ = 2.728$\times 10^{14}$~g~cm$^{-3}$.\\
(b) The mass densities at the center of compact stars in the unit of $\rho_0$ with typical gravitational masses, 1.4~$M_\odot$, 1.8~$M_\odot$, and the maximum mass $M_{\rm max}$ for each case of $\Sigma_{Kn}$ with $L$ = 65~MeV. The value of $M_{\rm max}$ is also shown in the unit of the solar mass $M_\odot$ for each case of $\Sigma_{Kn}$.}
\begin{center}
\noindent (a)~
\begin{tabular}{ c || c | c | c | c}
\hline
$\Sigma_{Kn}~ (\rm MeV) $  & $\rho^c(\rm DU)/\rho_0$ & $\rho^c(\Lambda)/\rho_0$  &   $\rho^c(\Xi^-)/\rho_0$  & $\rho^c(K^-)/\rho_0$ \\ \hline\hline
 0~(PYM)      &  2.49     & 2.82   & 3.97        & $-$       \\
 300      &  2.49   & 2.82         & 3.97       & 4.72       \\
 400      &  2.49     & 2.82            & 7.05         & 3.43       \\ \hline
\hline
\end{tabular} \\
\vspace{0.5cm}~
\noindent (b)~
\begin{tabular}{ c || c | c | c | c}
\hline
$\Sigma_{Kn}~ (\rm MeV) $ & $\rho_{\rm center}(1.4~M_\odot)/\rho_0$ &  $ \rho_{\rm center}(1.8~M_\odot)/\rho_0$   & $\rho_{\rm center}(M_{\rm max})/\rho_0$   & $M_{\rm max}/M_\odot$ \\ \hline\hline
 0~(PYM)      &  2.64     & 3.72    & 9.46       & 2.141      \\
300               &  2.64     & 3.72    & 9.10       & 2.124       \\
400               &  2.64     & 4.16   & 10.2       & 2.076       \\ \hline
\hline
\end{tabular}
\label{tab:onset}
\end{center}
\end{table*}

\begin{figure*}[t]

    \centering
    \includegraphics[width=0.8\linewidth]{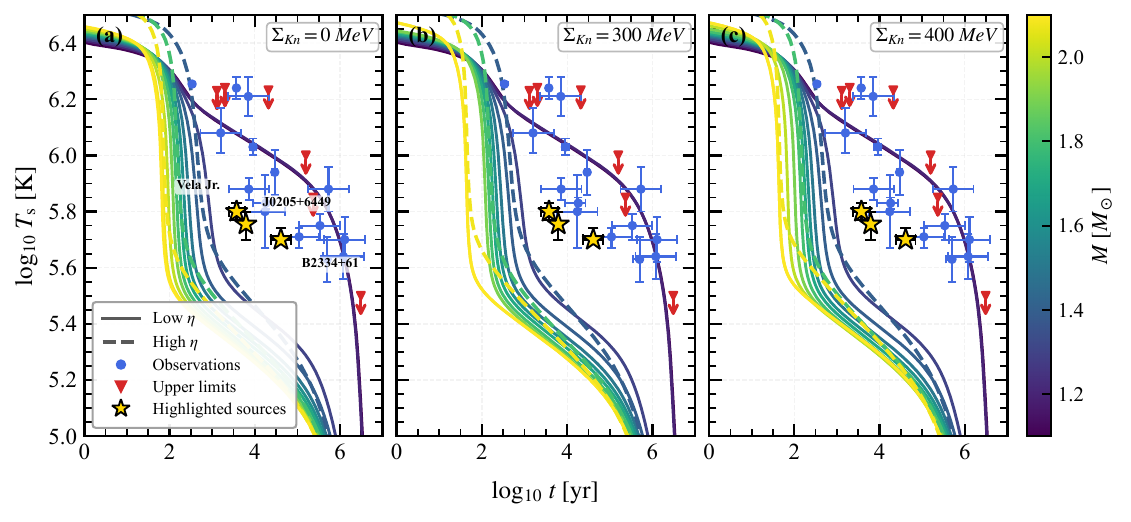}
    \caption{$\Sigma_{Kn}$ dependence of cooling curves without the inclusion of any SF model. The color contour represents the neutron-star mass from $1.1\,M_\odot$ and $2.1\,M_\odot$ with the grid of 0.02 $M_\odot$. The dashed lines correspond to cooling curves with fully accreted envelope (with only 1.4, 1.8, and 2.08 $M_\odot$), while the solid lines without an envelope. The error bars on the observations are taken from cooling observations of the NSs, with upper (and lower, if available) limits for age and temperature.
    }
    \label{fig:ccwosf}
\end{figure*}

\begin{figure}[t]
    \centering
    \includegraphics[width=\linewidth]{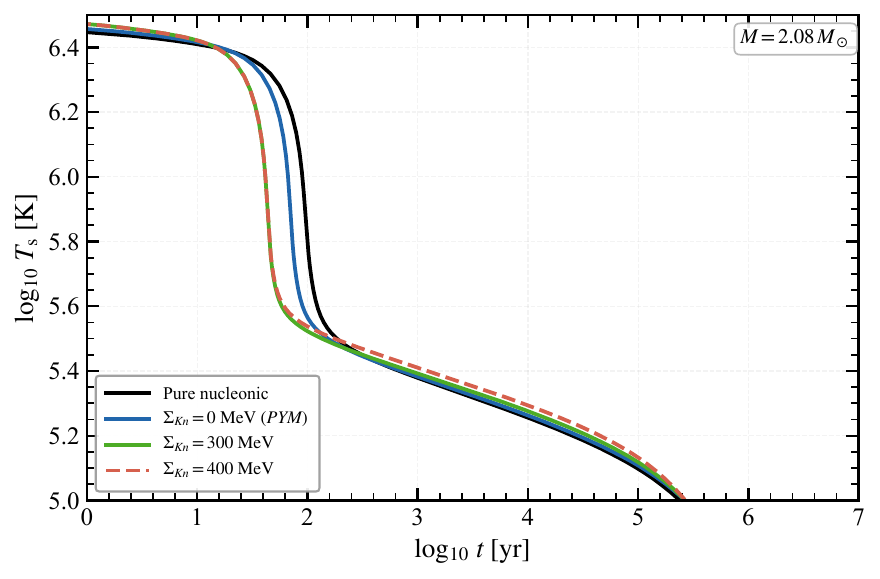}
    \caption{Cooling curves of $2.08~M_\odot$ NSs, with different EOSs: pure nucleonic, pure hyperonic, and hyperon-KC mixed matters. No envelope and superfluidity are considered.   
    }
    \label{fig:ccwosf2}
\end{figure}
\section{Cooling Calculation}
\label{sec:res}

To compute the thermal evolution of NSs, we use the publicly available \texttt{NSCool} code~\citep{Page2016}. This code solves the general relativistic energy balance and transport equations in spherically symmetric NSs with microphysics. Effects of hyperons ($\Lambda$, $\Xi^{-}$) and kaon ($K^{-}$) condensation are fully incorporated, such as the neutrino emission channel and specific heat of fermions\footnote{Note that the specific heat of kaons, which is proportional to $T^3$, is negligible because of
\begin{eqnarray}
\hspace*{-1cm}
    T\ll T_{{\rm BEC},K} = 
    4.15\times10^{10}~{\rm K}~\left(\frac{m_K}{m_K^*}\right)\left(\frac{Y_K}{0.01}\right)^{2/3}\left(\frac{n_{\rm B}}{n_0}\right)^{2/3},
\end{eqnarray}
where $T_{{\rm BEC},K}$ is the Bose temperature of kaons, which is much higher than the internal temperature of observed NSs ($T\lesssim10^9~{\rm K}$).}. 

\begin{figure*}[t]
    \centering
    \includegraphics[width=0.8\linewidth]{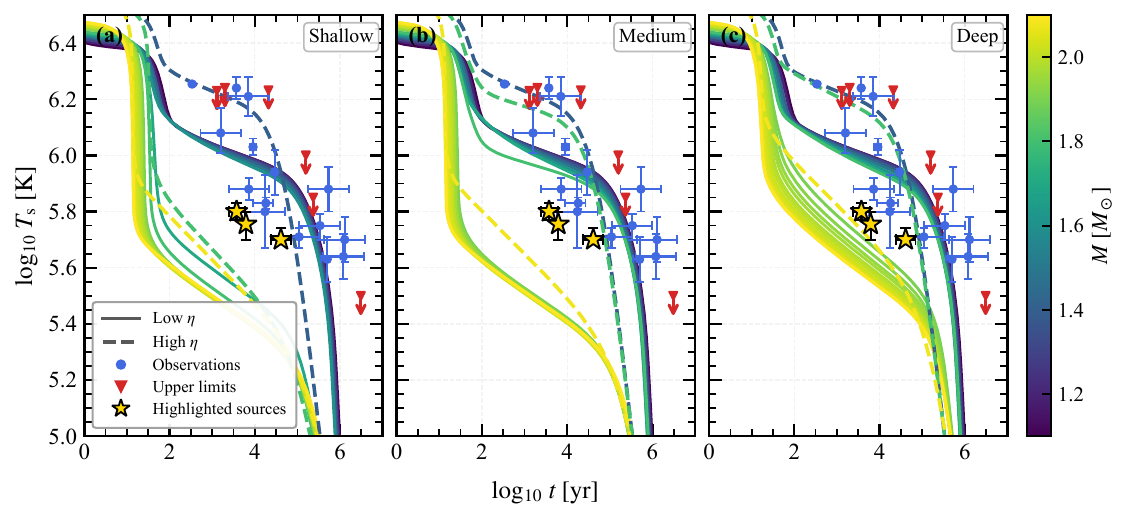}
    \caption{The same as Fig. \ref{fig:ccwosf}, but proton SF model dependence on cooling curves with $\Sigma_{Kn}$=300 MeV.
    }  \label{fig:ccwsf300}
    \centering
    \includegraphics[width=0.8\linewidth]{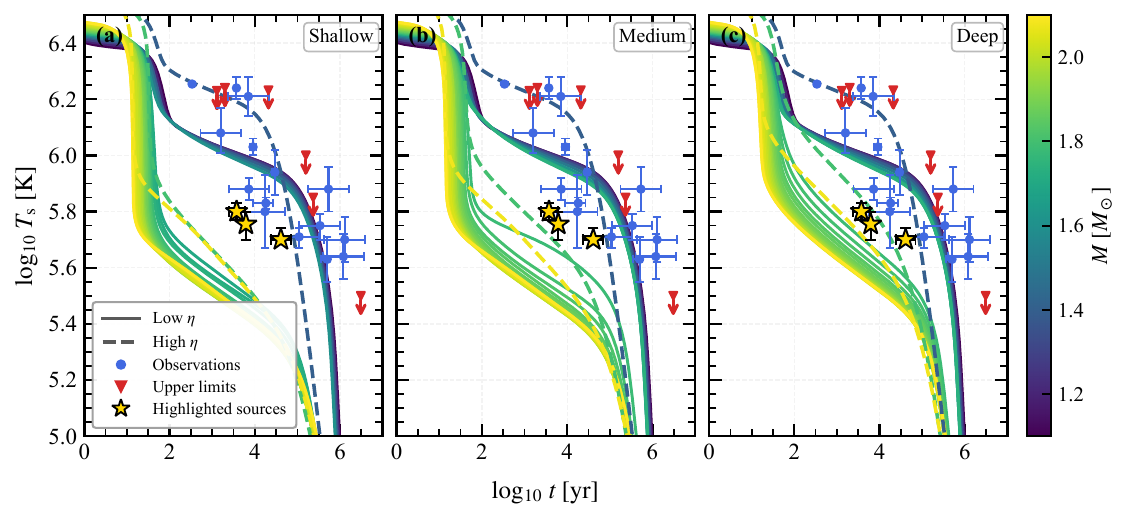}
    \caption{The same as \ref{fig:ccwsf300}, but for $\Sigma_{Kn}$=400 MeV. 
}
\label{fig:ccwsf400}
\end{figure*}

In addition to the EoS and SF models, there are remaining inputs to the cooling simulation of NS mass and envelope mass.  We compute cooling curves for stellar masses ranging from $1.1~M_\odot$ to $2.1~M_\odot$ in steps of $0.02~M_\odot$. We consider the widely-used stationary envelope model \cite{1997A&A...323..415P}, which has a free parameter of $\eta$ to characterize the mass of accreted envelope. Since the details of the envelope model are not our main focus, we consider two extreme cases of $\eta=10^{-30}$ (no envelope), and $\eta=10^{-7}$ (fully-accreted envelope). 

For observational constraints on the surface temperatures of isolated NSs, we mostly consider the same sources as in our previous work (Fig. 6 in \cite{Dohi2022}; see also Table 1 and Fig. 1 in \cite{2017IJMPE..2650015L}). In particular, we focus on three exceptionally cold NSs, namely CXOU J0852$-$4617 in Vela Jr., PSR J0205$+$6449 in 3C 58, and PSR B2334$+$61, whose thermal properties were recently reanalyzed by \cite{2024NatAs...8.1020M}. The unusually low thermal emission from these objects had already been discussed in earlier observational and cooling studies. For example, early Chandra observations placed a stringent upper limit on the surface temperature of PSR J0205$+$6449~\cite{Slane2002}, while XMM-Newton observations of PSR B2334$+$61 revealed predominantly thermal emission with a relatively low inferred surface temperature~\cite{McGowan2006}. For Vela Jr., its thermal emission and implications for NS cooling have also been investigated in previous studies~\cite{2002ApJ...580.1060K}. More recently, \cite{2026MNRAS.546ag126H} reanalyzed all available XMM-Newton and Chandra spectra of the Vela Jr. and further discussed its thermal properties in the context of NS cooling. Regarding the age of 3C58, we assume its lower and upper limits as 5400 and 7100 yrs, respectively~\cite{2006ApJ...645.1180B,2021ApJ...918L..33R} (see also \cite{2025AN....34670024F} for recent argument of historical age). We stress, however, that the positions of these sources on the cooling diagram are subject to uncertainties in their ages, distances, interstellar absorption, and spectral modeling. Our aim here is not to reanalyze the X-ray observations, but to examine whether the observed low temperatures can be accommodated within our cooling scenario.

Figure~\ref{fig:ccwosf} shows the cooling curves with the Y+K EoS under different assumptions of the $s$-wave  kaon--neutron scalar attraction (\(\Sigma_{Kn}\)), without including any SF effects. As evident from the figure, both values of \(\Sigma_{Kn}\) produce nearly identical timescales for the fast cooling phase, indicating only a minor quantitative difference between them. However, the critical mass separating the minimal and fast cooling regimes remains essentially unchanged at $M_{\rm crit}\approx1.3~M_\odot$, which coincides with the threshold mass \(M_{\rm DU}\) (see also Fig.~\ref{fig:M-rhoc}), suggesting that $np$DU dominates the cooling behavior in models exhibiting fast cooling. 

The impact of hyperons and KC on cooling curves is highlighted in Fig. \ref{fig:ccwosf2}. One can observe the occurrence of $\Lambda p$DU and KU, which shortens cooling timescale by a few 10 yrs. However, their changes on cooling curves are too small to be 
resolved from known temperature observations, due to the presence of $np$DU in lower-mass regions.




\begin{figure*}[t]
    \centering
    \includegraphics[width=0.9\linewidth]{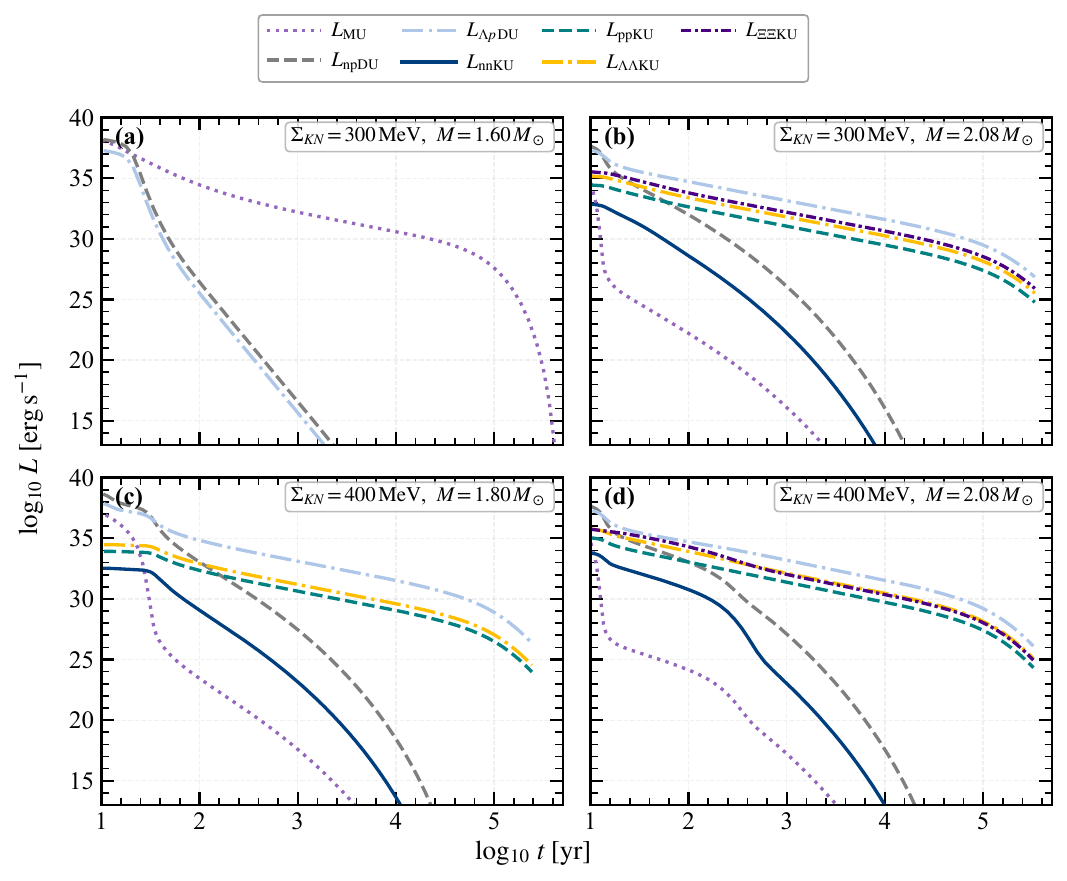}
    \caption{Luminosity evolution of fast neutrino cooling processes (and MU process) in case of \textit{shallow} proton SF model for $\Sigma_{Kn}$ = 300~MeV in upper panel and 400~MeV in lower panel. The left panel corresponds to the $1.6\,M_\odot$ stars, while the right panel $2.1\,M_\odot$ for $\Sigma_{Kn}$ = 300~MeV and $2.08\,M_\odot$ for  $\Sigma_{Kn}$ = 400~MeV. The curves represent contributions from MU process, direct Urca processes ($np$DU, $\Lambda p$DU) and kaon-induced Urca processes ($nn{\rm KU}$, $pp{\rm KU}$, $\Lambda\Lambda {\rm KU}$, $\Xi^- \Xi^- {\rm KU}$).}
    \label{fig:ems_shallow}
\end{figure*}

\begin{figure*}
    \centering
    \includegraphics[width=0.9\linewidth]{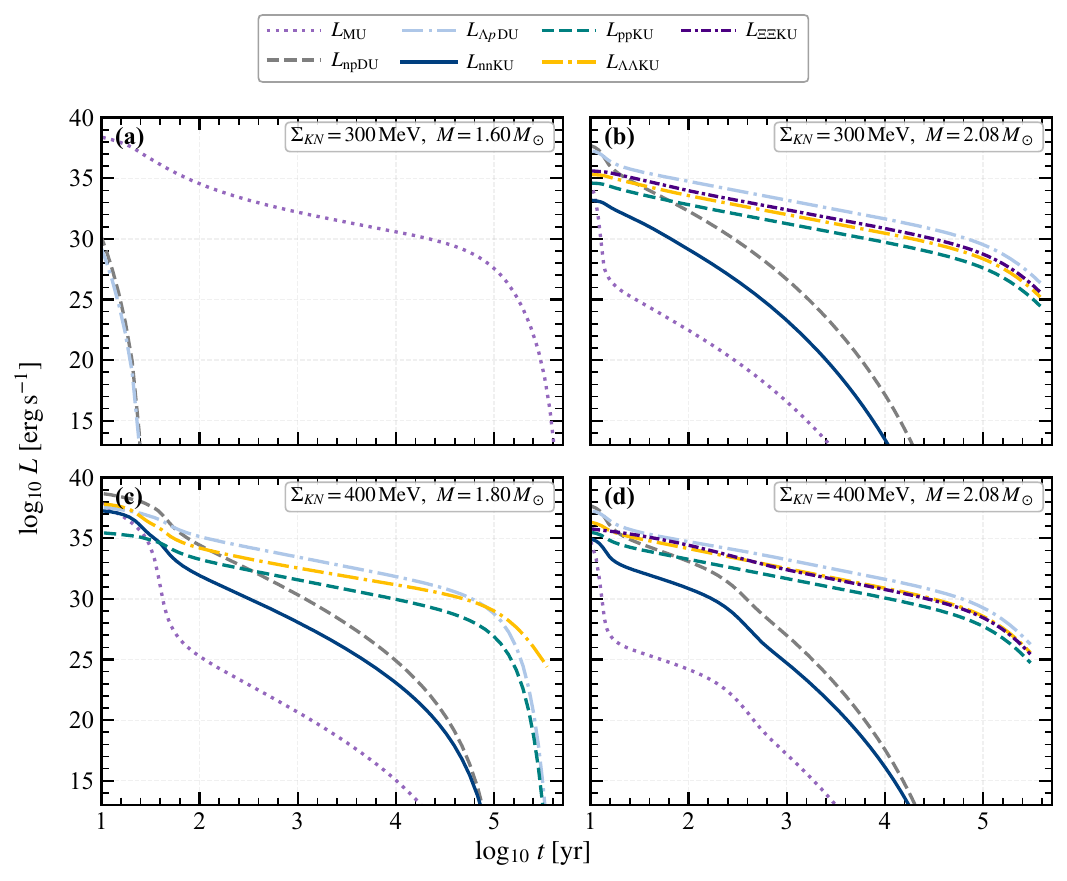}
    \caption{Same as Figure~\ref{fig:ems_shallow}, but for \textit{medium} proton SF model.}
    \label{fig:ems_medium}
\end{figure*}
\begin{figure*}
    \centering
    \includegraphics[width=0.9\linewidth]{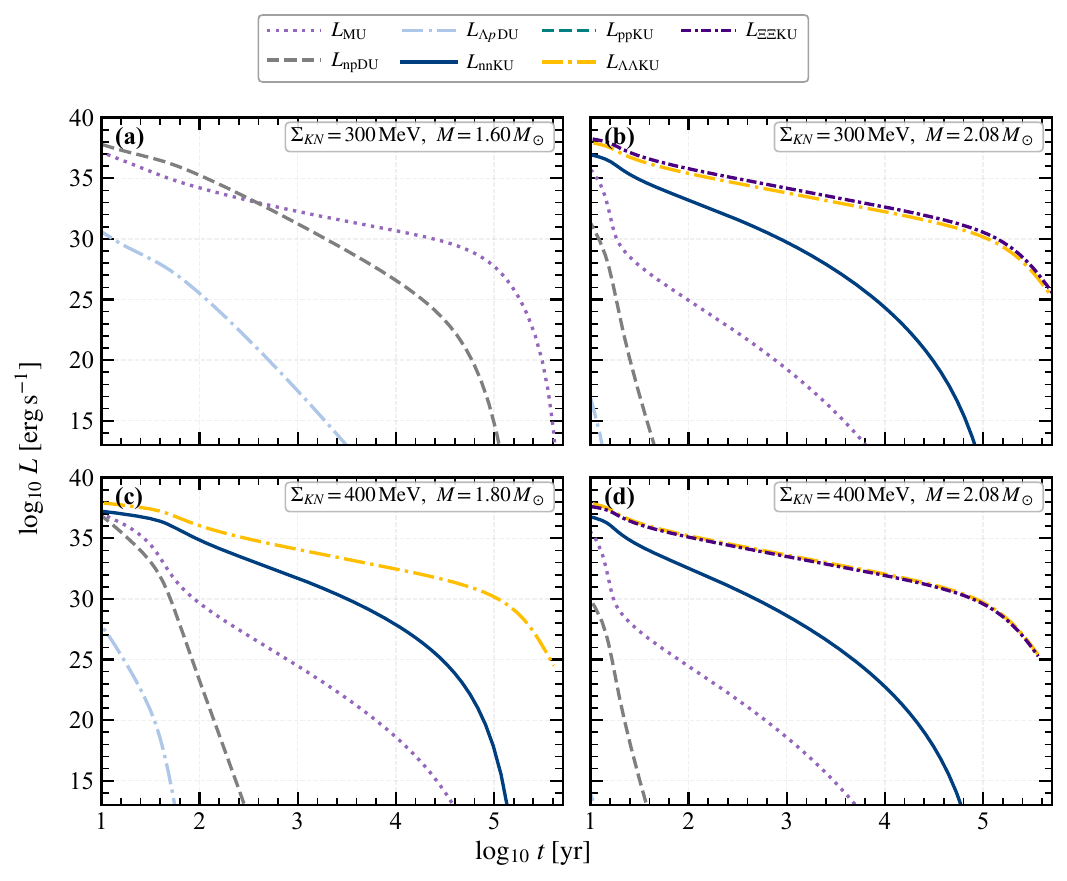}
    \caption{Same as Figure~\ref{fig:ems_shallow}, but for \textit{deep} proton SF model.}
    \label{fig:ems_deep}
\end{figure*}

Once the $np$DU occurs ($M_{\rm NS}\gtrsim1.3~M_\odot$), the surface temperature becomes too low to account for all observations, implying that cold NSs observed should be as light as 1.3 $M_\odot$. Because mass distribution is diverse with the average mass of $\sim1.4~M_\odot$~(e.g., \cite{2018MNRAS.478.1377A}), this scenario is unlikely. Thus, there should be missing physics, which is the very baryon SF that should exist in cold NSs: As the fast cooling is suppressed by baryon SF, as mentioned in Section \ref{sec:emissivity}E, the present result must become better for reproducing cold NS observations. 


\begin{figure*}[t]
    \centering
    \includegraphics[width=0.9\linewidth]{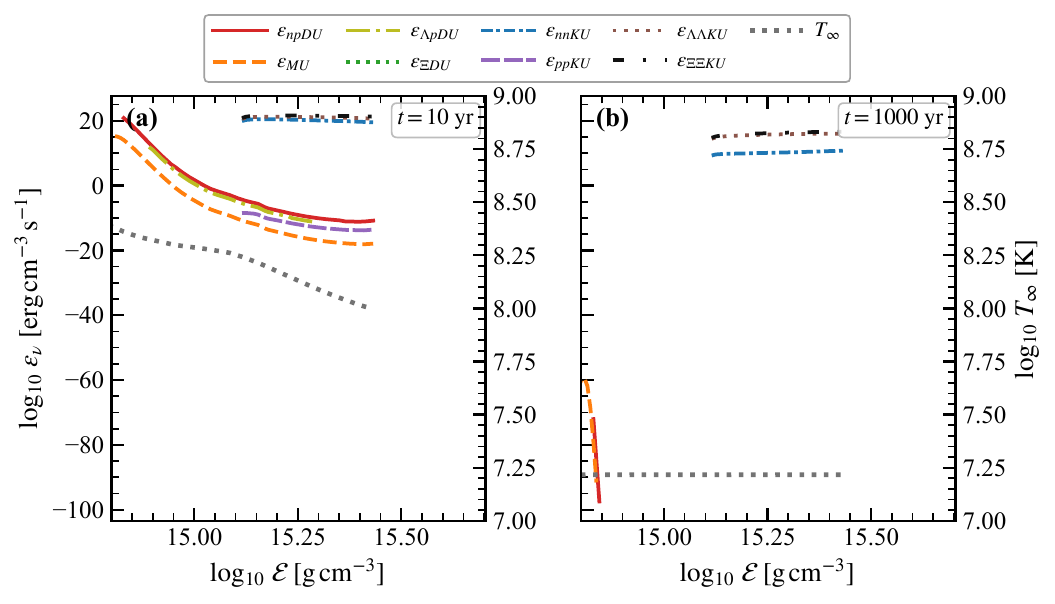}
    \caption{Two-time snapshot of density distribution of neutrino emissivity with $M_{\rm NS}=2.08~M_\odot$ and deep proton SF models for $\Sigma_{Kn}=300$\,MeV. The dotted gray line represents the corresponding temperature profile at an infinite observer.
\\
    }
    \label{fig:timestamp_300}
\end{figure*}


\begin{figure*}[t]
    \centering
    \includegraphics[width=0.9\linewidth]{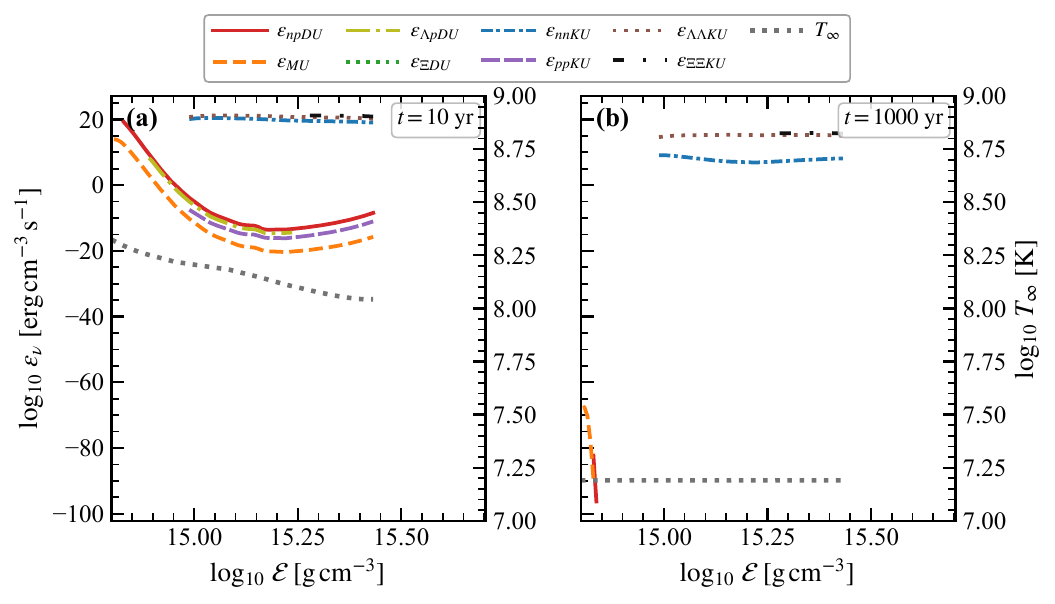}
    \caption{
    Same as Fig. \ref{fig:timestamp_300}, but for $\Sigma_{Kn}=400$\,MeV. 
    }
    \label{fig:timestamp_400}
\end{figure*}

Figs.~~\ref{fig:ccwsf300} and \ref{fig:ccwsf400} present cooling curves of SF NSs with $\Sigma_{Kn}=$300 and 400 MeV, respectively, focusing on proton SF. One can see that even ``shallow'' proton SF can significantly suppress rapid cooling of intermediate-mass NSs with $1.3\le M_{\rm NS}/M_\odot<M_{\rm crit}$, where the critical masses with shallow, medium and deep proton SF models are $M_{\rm crit}/M_\odot\approx$ 1.70(1.70),1.90(1.78), and 1.90(1.78) for $\Sigma_{Kn}=300(400)$ MeV, respectively. If the proton SF extends over high densities, i.e., in the case of deep model, cooling curves where the NS mass is slightly higher than $M_{\rm crit}$ pass through cold NS data such as 1.8 $M_\odot$ stars with $\Sigma_{Kn}=400~{\rm MeV}$, considering the uncertainties of envelope parameter $\eta$. This is owing to the suppression of DU processes while the unsuppression of KU processes, as we show later.

In the presence of nucleon SF, the strength of KC affects cooling curves, unlike the case without SF. Seen from above, one can see that the critical NSs with $\Sigma_{Kn}=400~{\rm MeV}$ become lower by a factor of 0.12 $M_\odot$ than those with $\Sigma_{Kn}=300~{\rm MeV}$, implying that stronger KC leads to faster cooling in heavy NSs.

Which kind of cooling process becomes dominant in the presence of proton SF? This feature is illustrated by the neutrino luminosity curves in Figs. \ref{fig:ems_shallow}--\ref{fig:ems_deep}. For 1.6 $M_\odot$ NSs, the slow cooling processes (e.g., neutron-branch MU process) are dominant at least for $t\gtrsim20~{\rm yr}$, which implies that both $np$DU and $\Lambda p$DU that occur are suppressed even in the case of the shallow proton SF model. For higher mass NSs, the behavior of cooling curves with shallow and medium proton SF models is similar in that $\Lambda p$DU is dominant, because of the absence of proton SF in high-density regions. This implies the possibility of seeing the signature of $\Lambda$ hyperons from cold NS observations, but not KC. This result is the same between different $\Sigma_{Kn}$ terms.

If the proton SF becomes so strong in high-density regions such as the deep model, both $np$DU and $\Lambda p$DU are suppressed even in high-density regions, as in Fig. \ref{fig:ems_deep}. In case of $\Sigma_{Kn}=300~{\rm MeV}$, we found that the hyperon KU becomes dominant with $2.08~M_\odot$ stars. Due to the difference of effective mass ratios in Fig. \ref{fig:meff}, $\Xi^-$$\Xi^-$KU becomes stronger than $\Lambda$$\Lambda$KU\footnote{Although our cooling calculation treats the emissivity of $\Lambda$$\Lambda$KU as that of $nn$ KU by changing $M^*_n/M_n\rightarrow M^*_\Lambda/M_\Lambda$, $\Lambda$$\Lambda$KU based on the \textit{direct} emissivity calculation actually has a minor role on NS cooling as compared with the $\Xi^-$$\Xi^-$KU  (see also Appendix A for the matrix elements of the $\Xi^-\Xi^-$KU and $\Lambda\Lambda$KU processes).}. Thus, from cold NS observations, one can specify the signature of $\Xi^-$-mixed KC if the proton SF is strong. 

In case of $\Sigma_{Kn}=400 {\rm MeV}$, KC appears below the onset density of $\Xi^-$ hyperons, which makes the $nn$KU dominant for $1.8~M_\odot$ stars. In high-mass NSs such as with $M_{\rm NS}=2.08~M_\odot$, $\Xi^-$$\Xi^-$KU becomes dominant as with the case of $\Sigma_{Kn}=300~{\rm MeV}$. 

Fig. \ref{fig:timestamp_300} and \ref{fig:timestamp_400} display evolution of internal temperature and neutrino emissivities with $2.08~M_\odot$ deep proton SF stars. At $t=10~{\rm yr}$, both $np$DU and $\Lambda p$DU appear in low-density NS core regions. At $t=10^3~{\rm yr}$, on the other hand, they are strongly suppressed due to the stronger proton SF pairing effects even at high densities. In this case, one can see that the $\Xi^-\Xi^-$KU, $\Lambda\Lambda$KU, and $nn$KU processes become dominant, whose situation continues up to the onset age of the photon cooling stage ($t\sim10^5~{\rm yr}$). Hence, this internal state with KC is principally visible from existing temperature observations with those ages (see Figs. \ref{fig:ccwsf300} and \ref{fig:ccwsf400}).

\section{Conclusion}
\label{sec:con}

Motivated by recent confirmation of cold isolated NSs, we investigated thermal evolution of NSs with mixed hyperons and KC for the first time, employing the latest Y+K EoS in line with recent mass and radius constraints. As expected from the density dependence of $Y_p$,  $np$DU cools non-superfluid NSs rapidly with $M_{\rm NS}\gtrsim1.3~M_\odot$, implying that no signature of strangeness can be seen from temperature observations. However, if the proton superconductivity is switched on, the $np$DU can be suppressed. In the shallow proton SF model, $\Lambda p$DU becomes dominant for NSs with $M_{\rm NS}\gtrsim1.8~M_\odot$, implying the signature of $\Lambda$ hyperons. Furthermore, if one considers the stronger proton SF in high-density regions, such as the deep model, not only $np$DU but also $\Lambda p$DU could be suppressed. Then, the dominant cooling processes become KU without protons. Thus, we conjecture that \textit{if the proton SF is strong ($T_{c,p}\sim10^{10}~{\rm K}$) in high-density regions, one can see the signature of KC from cold-NS observations.} 

This conjecture was drawn with the assumption of our Y+K EoS, but may be somewhat general because of the following reasons:
\begin{itemize}
    \item If one wants to get the stiff EoS with KC, since both symmetry energy and KC increase the proton fraction, $np$DU is likely to occur in low-mass NSs, as already mentioned.
    \item If the KC appears, the electron fraction becomes lower due to the equilibrium condition, which could prohibit DU processes associated with hyperons heavier than $\Lambda$. In the Y+K EoS, it corresponds to the $\Xi$-induced DU.
    \item The KC state, $|{\rm KC}\rangle$, can be regarded as the pathway state to the quark phase~\cite{2025Symm...17..270M} (i.e., its existence may be relevant to assistance of chiral restoration by affecting the density-dependence of chiral condensate, $\langle{\rm KC}|(\bar u u+\bar s s)|{\rm KC}\rangle$), which opens one of the rapid cooling processes, quark $\beta$ decay, depending on the color superconductive state (e.g., \cite{2013ApJ...765....1N}). However, the high-density quark phase tends to be the color-flavor-locked (CFL) phase, where quark $\beta$ decay is completely suppressed due to a very large gap $\sim100~{\rm MeV}$ (\cite{Alford08} and reference therein). 
\end{itemize}

Thus, the density dependence of the critical temperature of SF protons, $T_{c,p}(\rho)$ seems to be responsible for the observational visibility of KC. Therefore, our conclusion that kaon-induced Urca cooling becomes dominant should be interpreted as applying to the case where proton superconductivity extends into the density regime of kaon condensation, as represented by our phenomenological deep-pairing model. Although the realistic $T_{c,p}(\rho)$ value is highly uncertain, Xu et al. \cite{2018MNRAS.474.3576X} showed that KC leads to the obvious enhancement of the proton ${}^1{\rm S}_0$ SF, which may realize our conjecture. Takatsuka \& Tamagaki \cite{1995PThPh..94..457T} considered the proton ${}^3{\rm P}_2$ SF under KC, which is unlikely to appear due to a significant decrease of effective proton mass. As both studies are considered within the framework without hyperons, it is valuable to calculate $T_{c,p}$ under the Y+K phase, which is left for our future work.

Throughout this paper, we fix the neutron superfluid model and do not include hyperon superfluidity in order to isolate the role of proton superconductivity. Although different neutron pairing models may quantitatively modify the cooling curves through changes in neutron-related neutrino emission processes, neutron superfluidity alone cannot suppress the proton-dependent $\Lambda p$DU process. Therefore, uncertainties in the neutron pairing gap do not qualitatively alter our main conclusion that the kaon-induced Urca process becomes the dominant cooling mechanism only when proton superconductivity extends into the high-density core.
In terms of ${}^3{\rm P}_2$ neutrons, weak SF with $T_{c,n}\lesssim10^9~{\rm K}$ may be plausible in terms of both observations of Cassiopeia A and theoretical calculations (e.g., \cite{2011PhRvL.106h1101P,2023MNRAS.518.2775S}). Based on these inferences, neutron-induced KU is the most probable candidate to satisfy our conjecture. On the $\Lambda$ SF, it has been shown in the analysis of the ``Nagara'' event that the $\Lambda$-$\Lambda$ pairing attraction may be weak ($\sim$ 1~MeV)~\cite{Takahashi_2001,Ahn_2013} (but see also \cite{2019PTEP.2019b1D02E}). 
Although this is not relevant to the $\Lambda$$\Lambda$KU due to its minor role on the emissivity as compared with other fast cooling processes, the efficiency of $\Lambda p$DU would be kept to be high enough to dominate cooling curves of intermediate-mass NSs (see also \cite{2006PThPh.115..355T}) unless proton SF is too strong.

Once the $\Xi^-$$\Xi^-$ KU is allowed to occur, 
it could be suppressed by
$\Xi^-$ SF.
Tu \cite{2022PhRvC.106b5806T} calculated the hyperon pairing gap with several EoSs. He showed that $\Xi^-\Xi^-$ gap is large enough to suppress the $\Xi^-$-involved rapid cooling process in most density regions, but except for central density regions of very massive NSs; $T_{c,\Xi^-}$ in very high-density regions strongly depends on the properties of the pairing force. Further studies of hyperon gaps in high-density regions, therefore, could probe the possibility that the hyperon KU is visible from cold NS observations. 


This time, we focus on isolated NSs, but cold accreting neutron stars are also another site to probe fast neutrino cooling. In particular, quiescent luminosities of some soft X-ray transients such as SAX J1808.4$-$3658 and 1H 1905$+$000 are too low to be explained by minimal cooling scenario (\cite{Beznogov_2014} and reference therin). In the context of meson condensation, it is argued that the meson-induced Urca cooling is not likely to reproduce them due to its weakness compared to $np$DU and $\Lambda p$DU~\cite{2004ARA&A..42..169Y}. This does not hold in the case of pion condensation: Liu et al. \cite{Liu_2021} showed that cooling models that consistently treat physical properties of pion condensation~\cite{1993PThPS.112..221M} account for them as well as mass-radius observations, even a situation to allow the only pion Urca process among fast cooling processes. To check the extensibility of our conjecture on KC, it would be worth investigating thermal evolution of accreting NSs with Y+K EoS. 



\begin{acknowledgments}
The authors thank S. Tsuruta, T.~Tatsumi, T.~Maruyama, H.~Sotani, and N.~Yasutake for discussion and interest in this work. This work is supported by JSPS KAKENHI Grant Numbers JP25K17403(A.D.) and JP26K07094(T.N). T.M. appreciates the financial support from Chiba Institute of Technology. T.N. wishes to acknowledge the support from the Discretionary Budget of the President of Kurume Institute of Technology.
\end{acknowledgments}

\appendix
\section{Emissivity formula for Kaon-induced Urca process}

We recapitulate the expression of the neutrino emissivity for KU. Here only the nucleon-associated process, (NN-KU) \begin{subequations}\label{eq:KU}
\begin{eqnarray}
&& N + \langle K^-\rangle\rightarrow N + l +\bar\nu_l \qquad \ {\rm (a)} \ , \label{eq:KUa} \\
&& N + l \rightarrow N + \langle K^- \rangle + \nu_l  \qquad \ {\rm (b)} \label{eq:KUb}
 \end{eqnarray}
 \end{subequations}
(for $N=p, n$) is considered.  The emissivity for (a), $\epsilon_{\nu,NN{\rm KU}}^0$(a) without the effect of superfluidity denoted by the superscript ``0'' is given by
\begin{widetext}
\begin{equation}
\epsilon_{\nu,NN{\rm KU}}^0{\rm (a)}=\frac{V^3}{(2\pi)^{12}}\int d^3 p_N \int d^3 p_{N'} \int d^3 p_e 
\int d^3 p_{\bar\nu_e} E_{\bar\nu_e} W_{\rm fi}\cdot f_N({\bf p}_N)[1-f_{N'}({\bf p}_{N'})][1-f_e({\bf p}_e)]
 \ , 
\label{eq:emisKU-a}
\end{equation}
\end{widetext}
where $V$ is the normalization volume, $E_{\bar\nu_e}$ the (anti) neutrino energy, $W_{fi}$ the transition rate, and $f_N({\bf p}_N)$ $[=1/(1+\exp[(E({\bf p}_N)-\mu_N)/T]) ]$,  $f_e({\bf p}_e)$ are the Fermi distribution functions for the nucleon and electron, respectively\footnote{The Boltzmann constant $k_{\rm B}$ is put to be 1.}.
The transition rate $W_{fi}$ is written as
\begin{equation}
W_{\rm fi}=(2\pi)\delta(E_{\rm f}-E_{\rm i}-{\mu_K})\vert M\vert^2  \ , 
\label{eq:Wfi}
\end{equation}
where $\vert M\vert^2$ is the squared matrix element :
\begin{equation}
\vert M\vert^2 =\sum_{\rm spins} | \langle {\rm KC}; N, e^-, \bar\nu_e | H_W | {\rm KC}; N \rangle |^2 \ . 
\label{eq:M2}
\end{equation}

The weak Hamiltonian $H_W$ has the current $\times$ current form: 
\begin{equation}
H_W=\int d^3 r \frac{G_F}{\sqrt{2}} J_h^{\mu} l_\mu +{\rm h.c.} \ ,
\label{eq:wH}
\end{equation}
where $G_F$ is the Fermi coupling constant, $J_h^{\mu}$ the hadronic current, and $l_\mu$ is the leptonic current. 
$J_h^{\mu}$ is given by
\begin{equation}
J^\mu_{h}
=\cos\theta_c(V_{B,1+i2}^\mu-A_{B,1+i2}^\mu)+\sin\theta_c(V_{B, 4+i5}^\mu-A_{B, 4+i5}^\mu) \ , 
\label{eq:hc}
\end{equation}
where $\theta_c$ is the Cabibbo angle ($\sin\theta_c \simeq 0.23$), 
$V_{B,1\pm i2}^\mu\equiv V_{B,1}^\mu\pm i V_{B, 2}^\mu$ and $A_{B,1\pm i2}^\mu\equiv A_{B, 1}^\mu\pm i A_{B, 2}^\mu$ are the isospin-changing-strangeness-conserving vector and axial vector currents carried by baryons ($B$), respectively. Likewise $V_{B, 4\pm i5}^\mu\equiv V_{B, 4}^\mu\pm i V_{B, 5}^\mu$ and $A_{B, 4\pm i5}^\mu\equiv A_{B, 4}^\mu\pm i A_{B, 5}^\mu$ being the isospin-and strangeness-changing vector and axial vector currents carried by baryons $B$, respectively. 
The leptonic current for the electron in Eq.~(\ref{eq:wH}) is given as usual: 
 \begin{equation}
 l_\mu=\bar u_{\nu_e}\gamma_\mu(1-\gamma_5)u_e \ .
 \label{eq:lc}
 \end{equation}

The $V_{B, a}^\mu$ and $A_{B, a}^\mu$ ($a = 1\sim 8$) are given as~\cite{kubis_2006} 
\begin{widetext}
\begin{subequations}\label{eq:vacurrent}
\begin{eqnarray}
V_{B, a}^\mu&=&\frac{1}{4}{\rm Tr}\bar\Psi\gamma^\mu[u_{+, a}, \Psi]+\frac{F}{4}{\rm Tr}\bar\Psi\gamma^\mu\gamma^5[u_{-, a}, \Psi]+\frac{D}{4}{\rm Tr}\bar\Psi\gamma^\mu\gamma^5\lbrace u_{-, a}, \Psi\rbrace \ ,  
\label{eq:vcurrent}
\\
A_{B, a}^\mu&=&\frac{1}{4}{\rm Tr}\bar\Psi\gamma^\mu[u_{-, a}, \Psi]+\frac{F}{4}{\rm Tr}\bar\Psi\gamma^\mu\gamma^5[u_{+, a}, \Psi]+\frac{D}{4}{\rm Tr}\bar\Psi\gamma^\mu\gamma^5\lbrace u_{+, a}, \Psi\rbrace \ , 
\label{eq:acurrent}
\end{eqnarray}
\end{subequations}
\end{widetext}
where $\Psi$ is the baryon octet and 
\begin{equation}
u_{\pm, a}\equiv \xi^\dagger\lambda_a \xi\pm \xi\lambda_a\xi^\dagger
\label{eq:ua}
\end{equation}
with $\xi$ in the nonlinear representation of the classical kaon field by the use of Eq.~(\ref{eq:kfield}).

One can see from Eq.~(\ref{eq:Wfi}) that the kaon chemical potential $\mu_K$ is supplied to the energy conservation, $E_f-E_i-\mu_K=0$, which originates from the phase factor $\exp(-i\mu_K t)$ appearing in the hadronic current in the presence of KC in Eq.~(\ref{eq:vacurrent}). 

After calculating the transition matrix element by way of Eq.~(\ref{eq:hc}) and phase-space integral, one obtains
\begin{equation}
\epsilon_{\nu,NN{\rm KU}}^0{\rm (a)}=\frac{G_F^2}{16}\frac{457}{5040}\pi M_N^{\ast 2}\mu_e
\overline{\Big(H^{\mu\nu}L_{\mu\nu}\Big)}_{NN}  T^6  \ ,
\label{eq:emis-N-KU}
\end{equation}
where  $T$ is the temperature. 

In Eq.~(\ref{eq:emis-N-KU}),
{$\overline{\Big(H^{\mu\nu}L_{\mu\nu}\Big)}_{NN}$ is the angle-averaged reduced squared matrix elements with $H^{\mu\nu}$ being the hadronic tensor and $L^{\mu\nu}$ the leptonic tensor. For $N = p,n$ one obtains
\begin{subequations}\label{eq:aHL}
\begin{eqnarray}
\overline{\Big(H^{\mu\nu}L_{\mu\nu}\Big)}_{pp}&=&
4 (1+3F^2) \sin^2\theta_C  \sin^2\theta \ , 
\label{eq:aHL-1}  \\
\overline{\Big(H^{\mu\nu}L_{\mu\nu}\Big)}_{nn}&=&
\big\lbrace 1+3(D-F)^2\big\rbrace \cr
&\times&\sin^2\theta_C  \sin^2\theta  \ , 
\label{eq:aHL-2} 
 \end{eqnarray}
 \end{subequations}
where $D$ and $F$ are taken to be 0.81 and 0.44, respectively, with $g_A$ (=$D+F$=1.25) being the axial-vector coupling strength~\cite{1994NuPhA.571..758F}.

In the Y+K phase, the kaon-induced Urca process for hyperons (abbreviated to $YY$KU), where the nucleon $N$ in Eq.~(\ref{eq:KU}) is replaced by $Y$ (=$\Lambda, \Xi^-$), may become possible. 
For the $\Xi^-\Xi^-$KU process, $\Xi^- + \langle K^- \rangle \rightarrow \Xi^- + l +\bar\nu_l$, $\Xi^- +l \rightarrow \Xi^- + \langle K^- \rangle + \nu_l$, the transition matrix element is the same as that for the $pp$KU process [Eq.~(\ref{eq:aHL-1})]. 
For the $\Lambda\Lambda$KU process, $\Lambda + \langle K^- \rangle \rightarrow \Lambda + l +\bar\nu_l$, $\Lambda +l \rightarrow \Lambda + \langle K^- \rangle + \nu_l$, the transition matrix element yields $3D^2\sin^2\theta_c\sin^2\theta$.
The luminosity of these $YY$KU processes are given by not only the transition matrix element but also the available phase space, which depends upon the relevant particle fractions, shown in Fig.2. See also the right panels in Figs.~\ref{fig:ems_shallow},\ref{fig:ems_medium},\ref{fig:ems_deep} in the case of $\Sigma_{Kn}$ = 300~MeV and 400~MeV.

There are other neutrino emission processes induced by KC. For instance, ``strangeness-conserving'' $n\Lambda$KU processes,
 $n + \langle K^- \rangle \rightarrow \Lambda + l +\bar\nu_l$, $\Lambda +l \rightarrow n + \langle K^- \rangle + \nu_l$, may be possible as a unique process to KC ($\theta > 0$). The matrix element of this process is given by \break
 $6\cos^2\theta_c\sin^2(\theta/2)\lbrace 1+3(F+D/3)^2\rbrace$  without Cabibbo-suppression unlike the $NN$KU and $YY$KU processes. Therefore, the emissivity of the $n\Lambda$KU process may become as large as the one for the DU process for a fully developed KC case. Throughout this paper, however, we take into account only the $NN$KU and $YY$KU processes for kaon-induced neutrino emissions, and consider qualitative effects of KC on cooling behaviour of compact stars.


\bibliography{ref}{}
\bibliographystyle{apsrev4-2}

\end{document}